\documentclass[%
 aip,
 jcp
 amsmath,amssymb,
 reprint,%
]{revtex4-2}

\usepackage{graphicx}
\usepackage{dcolumn}
\usepackage{bm}

\usepackage[utf8]{inputenc}
\usepackage[T1]{fontenc}
\usepackage{mathptmx}
\usepackage{etoolbox}
\usepackage{color}
\usepackage{float} 
\usepackage{amsmath}
\usepackage{xcolor}
\usepackage{booktabs}
\usepackage[hidelinks]{hyperref}
\usepackage{cleveref}

\crefname{equation}{Eq.}{Eqs.}
\crefname{figure}{Fig.}{Figs.}
\crefname{table}{Table}{Tables}

\newcommand{\Citation}[1]{\textcolor{red}{{\footnotesize \tt Citation required}}}

\makeatletter
\def\@email#1#2{%
 \endgroup
 \patchcmd{\titleblock@produce}
  {\frontmatter@RRAPformat}
  {\frontmatter@RRAPformat{\produce@RRAP{*#1\href{mailto:#2}{#2}}}\frontmatter@RRAPformat}
  {}{}
}%
\makeatother
\begin{document}

\title{A Haldane-Anderson Hamiltonian Model for Hyperthermal Hydrogen Scattering from a Semiconductor Surface}
\author{Xuexun Lu}
\affiliation{Department of Chemistry, University of Warwick, Gibbet Hill Road, CV4 7AL, Coventry, UK}
\author{Nils Hertl}%
\affiliation{Department of Chemistry, University of Warwick, Gibbet Hill Road, CV4 7AL, Coventry, UK}%
\affiliation{Department of Physics, University of Warwick, Gibbet Hill Road, CV4 7AL, Coventry, UK}
\author{Sara Oregioni}
\affiliation{Department of Chemistry, Universit{\`a} degli Studi di Milano, Via Golgi 19, 20133 Milano, Italy}
 \author{Riley Preston}
  \affiliation{Institute of Physics, University of Freiburg, 79104 Freiburg, Germany}
\author{Samuel L. Rudge}
 \affiliation{Institute of Physics, University of Freiburg, 79104 Freiburg, Germany}
 \author{Michael Thoss}
 \affiliation{Institute of Physics, University of Freiburg, 79104 Freiburg, Germany}
\author{Rocco Martinazzo}
\affiliation{Department of Chemistry, Universit{\`a} degli Studi di Milano, Via Golgi 19, 20133 Milano, Italy}
\affiliation{Instituto di Scienze e Tecnologie Molecolari, CNR, via Golgi 19, 20133 Milano, Italy}
\author{Reinhard J. Maurer$^*$}
\email{r.maurer@warwick.ac.uk}
\affiliation{Department of Chemistry, University of Warwick, Gibbet Hill Road, CV4 7AL, Coventry, UK}%
\affiliation{Department of Physics, University of Warwick, Gibbet Hill Road, CV4 7AL, Coventry, UK}
\affiliation{ University of Vienna, Faculty of Physics, Kolingasse 14{--}16, 1090 Vienna, Austria}

\date{\today}

\begin{abstract}
Collisions of atoms and molecules with metal surfaces create electronic excitations in the metal, leading to nonadiabatic energy dissipation, inelastic scattering, and sticking. Mixed quantum-classical molecular dynamics simulation methods, such as molecular dynamics with electronic friction, are able to capture nonadiabatic energy loss during dynamics at metal surfaces. Hydrogen atom scattering from semiconductors, on the other hand, exhibits strong adsorbate-surface energy transfer only when the projectile kinetic energy exceeds the bandgap of the substrate. Electronic friction fails to describe this effect. Here, we report a first-principles parameterization of a simple Haldane-Anderson Hamiltonian model of hydrogen atom gas-surface scattering on Ge(111)$c(2\times8)$, for which hyperthermal scattering experiments have been reported. We subsequently perform independent-electron surface hopping and Ehrenfest dynamics simulations on this model, and validate these results through numerically exact quantum-dynamical simulations using the hierarchical equation of motion approach. While mean-field dynamics yield weak nonadiabatic energy loss that is independent of the initial kinetic energy, independent electron surface hopping simulations qualitatively agree with the experimental observation that nonadiabatic energy dissipation only occurs if the initial kinetic energy exceeds the bandgap of the surface.
\end{abstract}

\maketitle

\section{Introduction}

Well-defined collision experiments of atoms or molecules with surfaces, complemented by computer simulations, have shaped our understanding of energy transfer between impinging particles and the surface of the substrate. \cite{rettner1996, muino, auerbach21, rahinov2024} Energy exchange during a collision event can occur by coupling of adsorbate motion to substrate phonons or electronic excitations, so-called electron-hole-pair (EHP) excitations.~\cite{rittmeyer2018} The magnitude of one or the other energy loss mechanism depends on the electronic properties of the surface and the projectile and their respective atomic masses. EHP excitations contribute significantly in the case of metal substrates where no electronic bandgap exists, and when projectiles exhibit low electron affinities.~\cite{rahinov2024, kroes21} 

To account for nonadiabatic energy loss that arises due to EHP excitations,  molecular simulation methods are required that go beyond the Born-Oppenheimer approximation{---}the cornerstone of molecular dynamics (MD) simulation. One aspect is the electronic nature of the gaseous particle: nonadiabaticity plays only a minor role in the scattering dynamics of hyperthermal molecules with an electronic closed-shell configuration such as H$_2$\cite{rettner1996, luntz2005, muzas2012, stark2025} or CO. \cite{wagner2017, meng2022, meng2024, rahinov2024}  However, NO scattering experiments from metal surfaces revealed strong nonadiabatic effects which results in an efficient vibrational energy relaxation or excitation.\cite{huang2000, shenvi09, golibrzuch2013} The same picture holds for atom–surface scattering: the scattering of noble-gas atoms from metal surfaces is governed by lattice excitation and can be semi-quantitatively explained with a simple collision model.\cite{rettner1996, grimmelmann80} It has been shown that the energy transfer in hydrogen atom scattering from metal surfaces is dominated by EHP excitations.\cite{buenermann15, Dorenkamp18, hertl22b} This insight was obtained using molecular dynamics with electronic friction (MDEF) simulations, where EHPs are treated as a weak perturbation to the dynamics in the form of a dissipative frictional force.\cite{tully95, li_wahnstrom92a}
When the metal surface is replaced with an insulating surface or the surface is decorated with adsorbates, nonadiabatic effects do not significantly contribute and phonon excitation becomes the prime energy dissipation channel.\cite{buenermann15, Dorenkamp18b, hertl21b, lecroart21, liebetrau2024}

Recent experiments for hydrogen scattering from Ge(111)$c(2\times8)$ indicate that hyperthermal scattering of hydrogen atoms leads to a bimodal energy loss profile.~\cite{kruger2023} One maximum occurs at small energy losses, regardless of the initial kinetic energy. This can be completely explained by energy loss due to lattice excitations.\cite{kruger2023, kruger2024b} The second peak, however, only appears when the initial kinetic energy of the impinging hydrogen atoms exceeds the electronic bandgap. The center of this second peak corresponds to an energy loss that is always larger than the bandgap of the substrate. MDEF simulations at the level of the local density friction approximation,\cite{li_wahnstrom92a, li_wahnstrom92b, juaristi08} an established approach in gas-surface dynamics to include weak nonadiabatic effects, were not able to reproduce this bimodal curve but instead were found to produce a single peak that sits between the low energy loss and the high energy loss features observed in experiment.\cite{kruger2023} It was therefore postulated that the second peak arises from the excitation of valence electrons into the conduction band. This hypothesis was further corroborated by hydrogen atom scattering experiments with varying temperatures ranging from 150\,K to 1100\,K.\cite{kruger2024} With increasing temperature, the center of the inelastic energy loss peak shifts to lower values and the measured energy loss spectrum bears more and more similarity with measurements on conventional metal surfaces. 

Hyperthermal hydrogen atom scattering on Ge(111)$c(2\times8)$ has further been characterized with electronic structure theory and computational simulations: Based on first principles calculations of the band structure of hydrogen on different types of Ge surface atoms on the Ge(111)$c(2\times8)$ surface, it was postulated that nonadiabatic transitions from the valence into the conduction band only occur on specific sites on the surface, namely, the Ge rest atom site.\cite{zhuMechanisticInsightsNonadiabatic2024} This was further corroborated with time-dependent Kohn-Sham and classical path approximation simulations of the kinetic energy loss along selected trajectories. A shortcoming of the classical path approximation is that the simulation trajectory cannot respond to nonadiabatic transitions. 

Several nonadiabatic mixed quantum-classical simulation methods have been proposed in the past and applied to dynamics at metal surfaces. This includes the mean-field Ehrenfest method,\cite{Li2005}  the independent electron surface hopping (IESH) method,\cite{shenvi09, Gardner2023} and the broadened classical master equation (BCME) approach.\cite{Dou2016a} The IESH method is based on mapping the many-electron system onto an independent many-electron system represented by a Newns-Anderson Hamiltonian (NAH). A variety of mixed-quantum classical molecular dynamics simulation techniques based on the NAH picture have recently been benchmarked for low-dimensional model systems, describing molecule and atom interactions with metal surfaces.\cite{loaiza2018,gardner2022,gardner2023b} Furthermore, scattering simulations incorporating nonadiabatic and nuclear quantum effects via the hierarchical equations of motion (HEOM) approach were conducted, which can provide numerically exact reference data.\cite{dan2023,preston2025,Shi25} In Ref.~\citenum{preston2025}, both IESH and BCME were found to closely agree with full quantum dynamics in their description of nonadiabatic effects during  scattering of hyperthermal diatomic molecules from metal surfaces. Recently, IESH simulations have been performed on high-dimensional NAH representation for CO scattering from Au(111) based on constrained density functional theory calculations and machine learning representations.\cite{meng2024, hong2025} The resulting NAH model captured energy dissipation due to phonon and EHP excitation, showing the utility of the NAH representation for gas surface dynamics.

Nonadiabatic MQCD simulations are less reported for hyperthermal scattering dynamics at semiconductors. While fewest switches surface hopping is an established technique to simulate nonadiabatic transitions in molecular photochemistry, it is less commonly applied in solids and surface systems.\cite{barbatti2011, mai2020} Many existing surface hopping studies in semiconductor materials focus on aspects of charge carrier mobility.\cite{carof2019, lei2022,runeson2024}  Kretchmer and Chan reported the evolution of a projectile's spin during the scattering process and found that the depolarization during the scattering process decreases with increasing bandgap size.\cite{kretchmer2018} However, in this study, the classical path approximation was imposed, which decouples the electronic dynamics from the nuclear trajectory. Many existing MQCD simulation methods require the substrate to be metallic or make the wide band limit approximation, which implies that the substrate has a constant density of states in the spectral range relevant to the coupling between the adsorbate state and the substrate states.\cite{mizielinski2005} This is especially true for NAH-based MDEF\cite{mizielinski2005,jin2019} and BCME. Furthermore, the scope of Ehrenfest dynamics has been analyzed by Loaiza and Izmaylov\cite{loaiza2018} via analytically solvable models with a particular focus on metallic systems with many but finite states involved. For semiconductors, this assumption naturally breaks down, and thus, MDEF and BCME are not applicable. 

\begin{figure}
    \centering
    \includegraphics[width=3in]{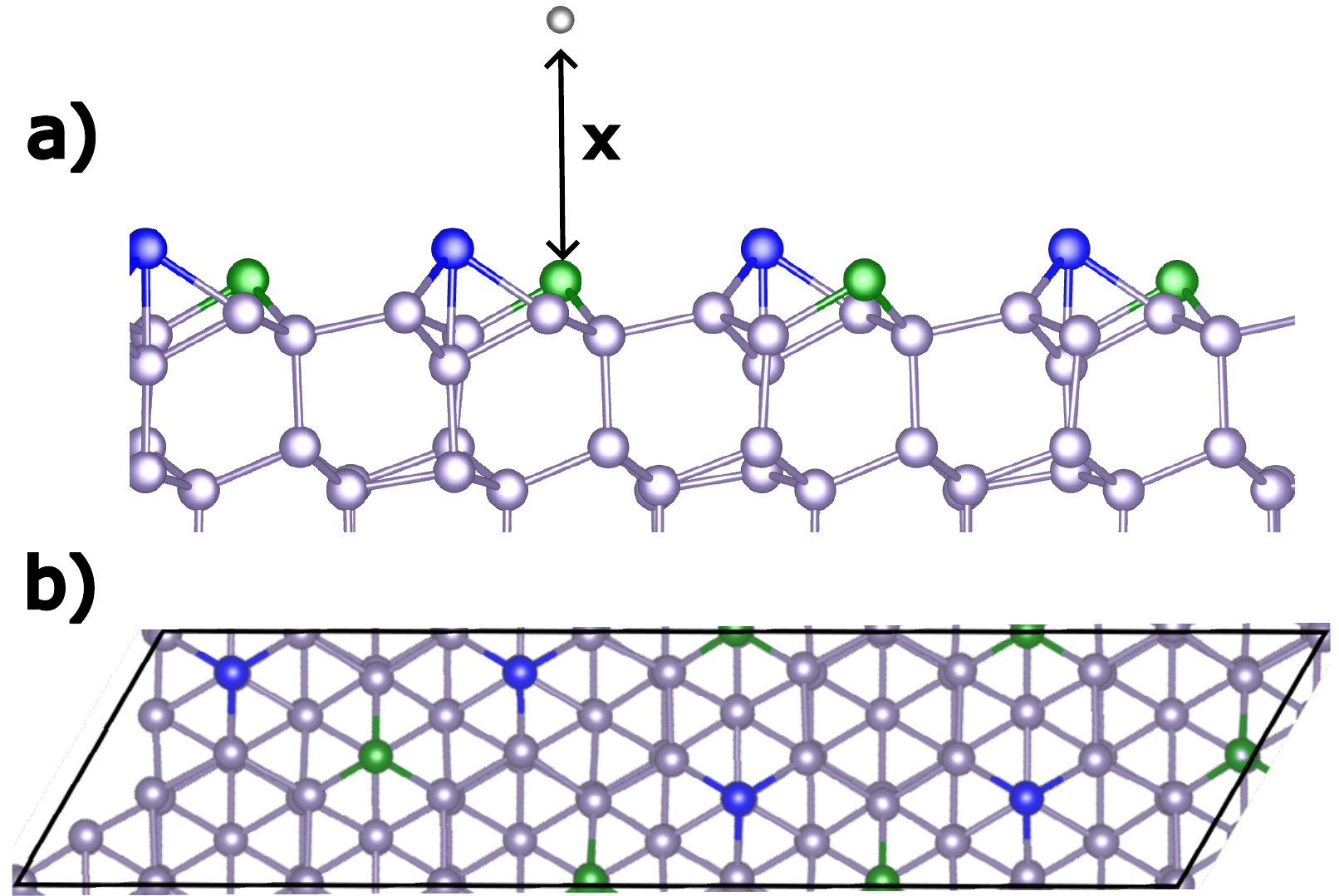}
    \caption{Atomic hydrogen (grey) at Ge(111)$c(2\times8)$ above a rest-atom (green). Adatoms are shown in blue and represent the second kind of reconstructed Ge atoms of this respective surface reconstruction. The surface is shown in side view along with the definition of the spatial model variable $x$ (panel a) and in orthographic top view (panel b).}
    \label{fig:HGe_structure}
\end{figure}

In this work, we develop a Haldane-Anderson model of a surface impurity in the presence of the electronic states of a semiconducting substrate. The bandgap in the electronic density of states of the substrate is realized through an adapted discretization scheme of the hybridization function, which effectively results in a similar model as previously proposed by Haldane and Anderson.\cite{haldane1976} The model is parametrized based on Density Functional Theory data to represent normal incidence scattering of hydrogen from the Ge rest atom of the Ge(111)$(2\times8)$ surface in one dimension. This choice is motivated by earlier work proposing that the electronic transitions are specific to this surface site.\cite{zhuMechanisticInsightsNonadiabatic2024} A representation of the surface adsorption geometry and the adsorption site can be seen in \cref{fig:HGe_structure}. Equipped with this model, we perform Ehrenfest and IESH simulations and find that Ehrenfest simulations do not provide results that are compatible with experimental observations. Ehrenfest predicts nonadiabatic energy loss at any initial kinetic energy---a shortcoming that we can trace back to the mean-field nature of the method. We find that IESH simulations of hydrogen atoms with initial kinetic energies smaller than the bandgap do not exhibit energy loss when scattered. However, once initial kinetic energies larger than the bandgap are applied, we obtain kinetic energy loss distributions that qualitatively match the highly inelastic scattering signal obtained in the experiment. The results from IESH simulations are validated against numerically exact quantum-dynamical simulations based on the HEOM approach, 
which incorporates all nonadiabatic and quantum nuclear effects, and we find quantitative agreement between the two. 
 
\section{Theory}

\subsection{Newns-Anderson Hamiltonian}

The Newns-Anderson Hamiltonian describes an electronic impurity in the presence of a continuum of electronic states, which can be used to represent an adsorbate electronic state in contact with $N$ electronic states of a metal surface.~\cite{Newns1969,Anderson1961} When parametrized with a nuclear degree of freedom, $x$, the Newns-Anderson Hamiltonian can be written as 
\begin{align}
    \hat{H}_\text{NA}(x, \hat{p}) = \frac{\hat{p}^2}{2m} + U_0(x) + \hat{H}_\mathrm{el}(x) ,
    \label{eq:Newns-Anderson-Hamiltonian}
\end{align}
where $x$ and $\hat{p}$ are the nuclear position and momentum operator of the particle with mass $m$. $U_0(x)$ is a potential that is independent of the electronic state. The third term $\hat{H}_{\text{el}}(x)$ is the electronic Hamiltonian parametrized by $x$ and has the following expression in the discretized second quantization notation: 

\begin{equation}
    \hat{H}_{\textrm{el}}(x) = h(x) \hat{d}^\dagger \hat{d} + \sum_{k=1}^{N}\epsilon_k \hat{c}_{k}^\dagger \hat{c}_{k} + \sum_{k=1}^{N}V_k(x)(\hat{d}^{\dagger}\hat{c}_{k} + \hat{c}_{k}^\dagger \hat{d}). \label{eq:Anderson-Hamiltonian-discrete}
\end{equation}
Here, $\hat{d}^\dagger(\hat{d})$ stand for the Fermionic creation (annihilation) operators for the electronic impurity state, whereas $\hat{c}_{k}^\dagger (\hat{c}_{k})$ are the Fermionic creation (annihilation) operators for an electron in the substrate electronic state $k$. If the impurity is occupied, the adsorbate state energy contributes to this Hamiltonian:
\begin{equation}
    h(x) = U_1(x) - U_0(x) .
    \label{eq:diabatic_difference}
\end{equation}
In the diabatic Newns-Anderson picture, $U_1(x)$ and $U_0(x)$ correspond to the charged and neutral state of the impurity, respectively. The first two terms in \cref{eq:Anderson-Hamiltonian-discrete} describe the energies of an impurity state and the $N$ continuum substrate states, while the last term represents their mutual coupling, with $V_k(x)$ being the coupling strength between substrate state $k$ and the impurity. This simple model can be readily extended to semiconductors by introducing a bandgap into the density of states of the substrate.\cite{haldane1976} 

\subsection{Mixed Quantum Classical Dynamics Methods}

We employ mixed quantum classical dynamics (MQCD) methods to incorporate coupling effects between the motion of the hydrogen atom and the electrons of the germanium surface upon collision. Electrons are propagated by the time-dependent Schr\"odinger equation, whereas the nuclei are treated as classical objects and thus follow Newton's equation of motion. Therefore, we neglect nuclear quantum effects. Given that the postulated electronic transition from the valence band of germanium into the conduction band is a strong nonadiabatic effect, the commonly-employed molecular dynamics with electronic friction (MDEF) method is not applicable as already shown by Kr\"uger \textit{et al.}.\cite{kruger2023}

\subsubsection{Independent Electron Surface Hopping Dynamics}

The independent electron surface hopping (IESH) approach is an extension of the fewest-switches surface hopping method\cite{tullyMolecularDynamicsElectronic1990b}  for nonadiabatic dynamics of adsorbates on metal surfaces.\cite{shenvi09} To enable surface hopping across $N+1$ electronic states, the correlated many-electron system is approximated as an independent $N_e$-electron system. As a consequence, the many-electron wavefunction $|\Psi\rangle$ can be formulated as a single Slater determinant 
\begin{equation}
    |\Psi\rangle =|\psi_1\cdots\psi_l\cdots\psi_{N_e}\rangle
\end{equation}
where $|\psi_l\rangle$ is the $l^{\text{th}}$ occupied single electron state and the electronic Hamiltonian can be represented by a sum of single particle Hamiltonians, represented in the form of the NAH:
\begin{equation}
    \hat{H}_\text{el}^1(x) = h(x)|d\rangle\langle d| + \sum_{k=1}^N \varepsilon_k |k\rangle\langle k| + \sum_{k=1}^N V_{k}(x)\left(|d\rangle\langle k| + |k\rangle\langle d|\right),
    \label{eq:single-electron-hamiltonian}
\end{equation}
where $|d\rangle$ and $|k\rangle$ are single electron states created by the action of operators $d^\dagger$ and $c_k^\dagger$, respectively.

The $N_e$ one-electron wavefunctions can be expanded in the basis of the adiabatic eigenstates of the single-electron Hamiltonian $\hat{H}_\text{el}^1(x)$
\begin{equation}
    \psi_l(x,t) = \sum_{k=1}^{N+1} c_k^l(t) \phi_k(x)
    \label{eq:single-electron_WF}
\end{equation}
with $c_k^l(t)$ being the time-dependent expansion coefficients and $\phi_k(x)$ the eigenstates of $\hat{H}_\text{el}^1(x)$. The expansion coefficients can be propagated by the time-dependent Schr\"odinger equation:
\begin{equation}
    i\hbar \dot{c}_k^l = \lambda_k(x)c_k^l -i\hbar \sum_{j\neq k} \frac{p}{m} d_{jk}(x)c_j^l ,
    \label{eq:TDSE}
\end{equation}
where $p$ stands for the nuclear momentum and $\lambda_k(x)$ is the eigenvalue for state $k$ and $d_{jk}(x)$ is the nonadiabatic coupling vector between state $j$ and $k$:
\begin{equation}
    d_{j k}(x)=\frac{\left(Q^{\dagger}(x) \frac{\partial \hat{H}_{\mathrm{el}}}{\partial x} Q(x)\right)_{j k}}{\lambda_j-\lambda_k}.
    \label{eq:na_couplings}
\end{equation}

Here, $Q(x)$ stands for a unitary matrix constituted by the eigenvectors of $ \hat{H}_\text{el}^1(x)$ which enables rotation from the diabatic into the adiabatic representation. The Hamiltonian that defines the classical nuclear propagation in IESH combines the kinetic energy of the nuclei, the electronic Hamiltonian in adiabatic representation, and the state-independent potential $U_0(x)$ and:
\begin{equation}
    H_\text{IESH}(x,p) = \frac{p^2}{2m} + U_0(x) + \sum_{k\in \bm{s}(t)} \lambda_k(x).
    \label{eq:IESH_Hamiltonian}
\end{equation}
The potential energy of $H_{\text{IESH}}$ comprises the state-independent term $U_0(x)$ and the sum of $N_e$ occupied eigenvalues of $ \hat{H}_\text{el}^1(x)$. The state vector $\bm{s}(t)$ lists the indices of the electronic states that are occupied at time $t$. $\bm{s}(t)$ connects the electron dynamics governed by \cref{eq:TDSE} and the nuclear dynamics governed by \cref{eq:IESH_Hamiltonian}. The state vector is subject to change due to surface hopping events that can occur at every time-step, which in turn changes the potential energy surface in $H_\text{IESH}$. Within IESH, only up to one electron is allowed to change its occupied state through a hop per time-step, which limits the maximum time step in regions of strong nonadiabatic coupling. The probability for a hop from state $k \rightarrow j $, which replaces index $k$ in $\bm{s}(t)$ with $j$, is evaluated by
\begin{equation}
    g_{k\rightarrow j} = \text{max}\left(\frac{B_{jk} \Delta t}{A_{kk}}, 0 \right),
\end{equation}
with
\begin{equation}
    B_{jk} = -2\text{Re}\left(A_{kj}^* \right) \frac{p}{m}d_{jk}.
\end{equation}
The quantity $A_{kj}$ is calculated from the many-body electronic wavefunction and the occupation vectors $|\bm{k}\rangle$ and $|\bm{j}\rangle$  via 
\begin{equation}
    A_{kj} = \langle \bm{k}| \Psi\rangle \langle\Psi|\bm{j}\rangle 
\end{equation}
The occupation vector $|\bm{k}\rangle$ and $|\bm{j}\rangle$ are eigenstates of the many-electron Hamiltonian and can be expressed as single Slater determinants with the underlying basis functions being the eigenfunctions of $ \hat{H}_\text{el}^1(x)$. Hence, the inner product of $|\bm{k}\rangle$ and $|\Psi\rangle$ is thus
\begin{equation}
     \langle \bm{k}| \Psi\rangle = \text{det}|\bm{S}|
\end{equation}
with the entries of the overlap matrix $\bm{S}$ being
\begin{equation}
    S_{ij} = c_{k_i}^{(j)},
\end{equation}
where $k_i$ is the single-electron state $k$ which is occupied by electron $i$ in the many-electron state $|\bm{k}\rangle$ and $\bm{c}^{j}$ is the vector of expansion coefficients of electron $j$.
The hopping probabilities are evaluated and compared to a uniform random number between 0 and 1. If a hopping event is selected to happen, the velocity is rescaled in the direction of the nonadiabatic coupling vector to ensure that the energy is conserved. If there is a sufficient amount of kinetic energy to overcome the jump in potential energy, the hop is allowed to occur and the vector $\bm{s}(t+\Delta t)$ is modified; otherwise, the hop is rejected and the electronic population of the total system remains unchanged. Upon a successful hop, the velocity of the particle is rescaled along the direction of the nonadiabatic coupling vector to ensure energy conservation.

\subsubsection{Mean-field Ehrenfest Dynamics}

In Ehrenfest dynamics (ED), the classical particle evolves  on the following time-dependent Hamiltonian:
\begin{equation}
    H_{\text{Ehr}}(x,p) = \frac{p^2}{2m} + U_0(x) + \langle \Psi(t) | \hat{H}_{\text{el}}(x) | \Psi(t) \rangle,
    \label{eq:Ehrenfest-Hamiltonian}
\end{equation}
where $\langle \Psi(t) | \hat{H}_{\text{el}}(x) | \Psi(t) \rangle$ is the time-dependent expectation value of the electronic Hamiltonian. The force acting on the nuclei in the simulation is therefore
\begin{align}
    F &= - \frac{\partial H_\text{Ehr}(x, p)}{\partial x} \\
    &= - \frac{\partial U_0(x)}{\partial x} - \frac{\partial \langle \Psi(t) | \hat{H}_{\text{el}}(x) | \Psi(t) \rangle}{\partial x}
    \label{eq:Ehrenfest-mean-force}
\end{align} 

When defining the many-electron Hamiltonian as an independent $N_e$-electron system, we can propagate the electronic wavefunction according to \cref{eq:TDSE}. We can further reexpress \cref{eq:Ehrenfest-Hamiltonian} in terms of the eigenstates of the Hamiltonian and the time-dependent coefficients:
\begin{equation}
    H_{\text{Ehr}}(x,p) = \frac{p^2}{2m} + U_0(x) + \sum_k |c_k(t)|^2 \lambda_k(x)
    \label{eq:Ehrenfest-Hamiltonian2}
\end{equation}
\Cref{eq:Ehrenfest-Hamiltonian2} closely follows the IESH Hamiltonian except that all eigenvalues contribute to the smoothly varying energy landscape according to their time-dependent electronic population. In the IESH Hamiltonian, all electronic states only contribute as fully occupied states, whereas in ED, electronic states contribute to the energy landscape according to their fractional population during the dynamics. Unlike in IESH, Ehrenfest trajectories are deterministic and thus entirely determined by their initial conditions.

\subsection{Hierarchical Equations of Motion}

The hierarchical equations of motion (HEOM) approach is a numerically exact, fully quantum method for modeling the dynamics of open quantum systems. As such, it can provide benchmarks for MQC methods. \cite{preston2025} Here, we give a brief overview of the method, leaving the more technical details to Refs.~\citenum{tanimura89,Tanimura2006,tanimura20,Haertle2015,jin07,jin08,schinabeck18,zheng09,yan14,wenderoth16,ye16}.

\begin{figure}[h]
\includegraphics[scale=0.6]{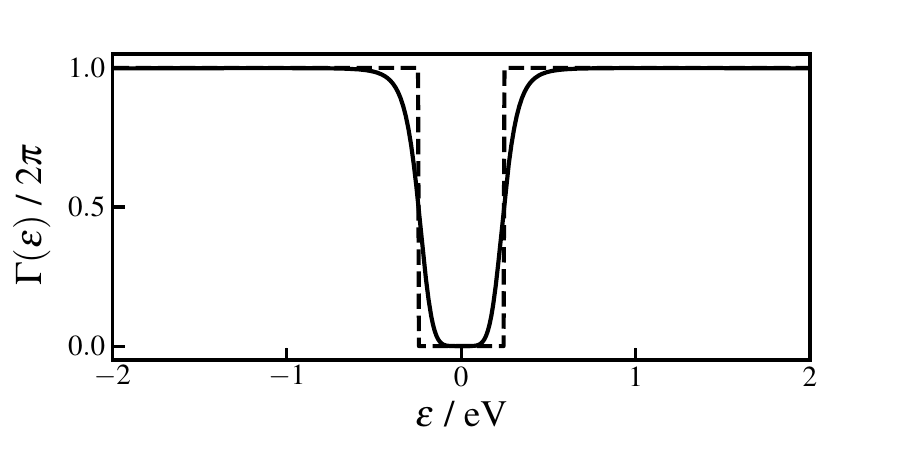}
\caption{Comparison of the molecule-surface couplings, $\Gamma(\varepsilon)$, used in the MQC (dashed line) and HEOM (solid line) methods. In the HEOM approach, the bandwidth is chosen as $W = 50$eV and the steepness parameter is $m = 3$, whereas the MQC calculations explicitly enforce the wide-band  and infinite-$m$ limit.}
\label{fig: HEOM_DOS}
\end{figure}

The surface is treated as a continuum of noninteracting electronic states held at local equilibrium with temperature $T$ and chemical potential $\mu$. The occupation function of the surface is given by 
\begin{equation}
f^{\sigma}(\epsilon) = \left(1 + e^{\sigma(\epsilon - \mu) / k_B T}\right)^{-1},
\label{eq: fermi-dirac}
\end{equation}
where $\sigma \in \{+,-\}$, with $\sigma = +$ for electrons and $\sigma=-$ for holes. The electronic interaction of the molecule with the surface is quantified by the spectral density:
\begin{align}
\tilde\Gamma(\epsilon,x) = \: & 2\pi \sum_{k} V_{k}^2(x) \delta(\epsilon - \epsilon_{k}).
\end{align}
We assume that it can be factorized into coordinate- and energy-dependent components,
\begin{align}
\tilde\Gamma(\epsilon,x) = V^2(x) \Gamma(\epsilon),
\end{align}
where the normalized strength of the electronic coupling to the surface is now quantified by $V(x)$. To model the band gap required for a semiconductor, we take $\Gamma (\epsilon)$ to have the form
\begin{equation}
\Gamma(\epsilon) = 2\pi \frac{W^2}{(W^2 + \epsilon^2)} \frac{\epsilon^{2m}}{(\epsilon^{2m} + (E_\text{gap}/2)^{2m})}.
\label{eq:HEOM_density}
\end{equation}
As shown in Fig.~\ref{fig: HEOM_DOS}, this has the functional form of a Lorentzian of width $W$ with a band gap of width $E_\text{gap}$ centered at $\epsilon = 0$. The parameter $m$ tunes the steepness of the band gap edge, such that for large $m$ it approaches the sharp band gap used in the MQC methods, also shown in Fig.~\ref{fig: HEOM_DOS} for $W = 50$\,eV and $m=3$.

The HEOM method propagates the reduced density matrix of the molecule, $\rho_{\text{mol}}$, obtained by integrating out the surface degrees of freedom. The influence of the molecule-surface coupling on the dynamics of the molecule arises via the bath-correlation functions,
\begin{equation}
C^\sigma_\text{el} (t) = \frac{1}{2\pi}\int^\infty_{-\infty}e^{i\sigma\epsilon t}\tilde{\Gamma} (\epsilon,x) f^{\sigma}(\epsilon)d\epsilon,
\label{BCF}
\end{equation}
which are decomposed into a sum over exponential functions for use in HEOM:
\begin{equation}
C^\sigma_\text{el} (t) \approx  V^2(x)\sum_{p=1}^{K/2}  \eta^\text{el}_{p\sigma} e^{-\gamma^\text{el}_{p\sigma} t}, \label{eq: bcf decomposition}
\end{equation}
This decomposition effectively maps the continuum of electronic states in the surface onto a finite set of $K$ virtual fermionic levels. Although several different pole decomposition schemes have previously been employed \cite{}, in this work we use the AAA decomposition scheme \cite{barycentric23,takahashi24} to decompose the occupation function in \cref{eq: fermi-dirac}. In contrast, the spectral function modeling the bandgap, given in \cref{eq:HEOM_density}, can be analytically decomposed in terms of $m+1$ unique poles. The sum over exponential functions in \cref{eq: bcf decomposition} runs over both sets of poles.

The influence of surface phonons can also be included in the HEOM framework by coupling to a phonon bath in an equivalent manner to the coupling to an electronic reservoir. \cite{yaling22} For the sake of simplicity, we model the phonon bath by a Lorentzian spectral density,
\begin{align}
\tilde J(\epsilon,x) = 2 \Lambda^{2}(x) \frac{\epsilon \Omega_{c}}{\epsilon^{2} + \Omega_{c}^{2}},
\end{align}
where $\Omega_c$ is the cutoff frequency and $\Lambda(x)$ is the coordinate-dependent coupling strength. $\Lambda (x)$ is taken to have the following phenomenological dependence on $x$,
\begin{equation}
    \Lambda (x) = \lambda e^{-\alpha x}.
\end{equation}
The correlation function for the phonon bath, 
\begin{equation}
C_\text{ph}(t) =  \frac{1}{2\pi} \int^{\infty}_{-\infty} d\epsilon \: J(\epsilon)[2 f_{B}(\epsilon) + 1] e^{i \epsilon t},
\end{equation}
is similarly decomposed as a power series of exponential functions,
\begin{equation}
C_\text{ph}(t) \approx \Lambda^{2}(x)\sum_{p = 1 }^{L} \eta^{\text{ph.}}_{p}e^{-\gamma^{\text{ph.}}_{p} t},
\label{phonon_pole_decomp}
\end{equation}
so that the interaction with the continuum of phonon modes in the surface is mapped onto $L$ effective virtual phonon modes.

The influence of the phonon and electronic bath-correlation functions on the molecule can be expressed with the Feynman-Vernon influence functional approach. Given the self-similarity of the summands of the bath-correlation functions in \cref{eq: bcf decomposition} and \cref{phonon_pole_decomp} with respect to a time derivative, this can then be written as a hierarchical set of coupled equations of motion of the form, \cite{yaling22,tanimura20,tanimura89,jin07,jin08,schinabeck18,hartle15,zheng09,yan14,wenderoth16,ye16}
\begin{widetext}
\begin{multline}
\frac{d\rho^{\mathbf n, \mathbf m}(t)}{dt} = 
\
-i\left[H_\text{mol}, \rho ^{\mathbf n, \mathbf m}(t) \right]
\
- \sum_{k=1}^K n_k \gamma^\text{el}_{p_k \sigma_k }\rho^{\mathbf n, \mathbf m}(t)
\
+ \sum_{l=1}^L m_l \gamma^\text{ph}_{p_l}\rho^{\mathbf n, \mathbf m}(t)
\\
+i\sum_{l=1}^K\sum_{k=1}^K \delta_{\bar\sigma_k, \bar\sigma_l} \delta_{p_k, p_l} (-1)^{\sum_{j<k} n_j} \sqrt{1-n_k} V_{i_k} \left(d_{i_k}^{\bar\sigma_k} \rho^{\mathbf{n+1}_l, \mathbf m}(t) + (-1)^{|\mathbf{n}|+1}\rho^{\mathbf{n+1}_l, \mathbf m}(t)d^{\bar\sigma_k}_{i_k}\right)
\
\\
\
+i\sum_{k=1}^K (-1)^{\sum_{j<k} n_j} \sqrt{n_k} V_{i_k} \left(\eta^\text{el}_{p_k \sigma_k} d_{i_k}^{\sigma_k} \rho^{\mathbf{n-1}_k, \mathbf m}(t) - (-1)^{|\mathbf{n}|-1}\rho^{\mathbf{n-1}_k, \mathbf m}(t)\eta^\text{el*}_{p_k \bar\sigma_k}d^{\sigma_k}_{i_k}\right)
\
\\
\
+i\sum_{l=1}^L\sqrt{m_l + 1} (\Lambda(x) \rho^{\mathbf{n}, \mathbf m + 1_l}(t) - \rho^{\mathbf{n}, \mathbf m + 1_l}(t) \Lambda(x)) 
\
+i\sum_{l=1}^L\sqrt{m_l}(\eta_{p_l}^\text{ph} \Lambda(x)  \rho^{\mathbf{n}, \mathbf m - 1_l}(t) - \rho^{\mathbf{n}, \mathbf m - 1_l}(t) \eta_{p_l}^\text{ph*} \Lambda(x) ).
\label{HEOM}
\end{multline}
\end{widetext}
Here, $\mathbf{n} = [n_1,n_2,...,n_K]$ and $\mathbf{m} = [m_1,m_2,...,m_L]$ are the vectors of occupations of the virtual fermionic states and virtual phonon bath modes, respectively. The reduced density matrix of the scattering molecule is obtained by setting the occupations of all virtual states to zero, so that $\rho_\text{mol} = \rho^{([0,0,...,0],[0,0,...,0])}$. Each unique configuration of occupations of the virtual states with non-zero norm corresponds to an auxiliary density operator (ADO), $\rho^{\mathbf{n},\mathbf{m}}$, where the norm of the virtual fermionic states is $\rvert \mathbf n \rvert = \sum_{k=1}^K n_k$ and the norm of the virtual phonon modes is $\rvert \mathbf m \rvert = \sum_{l=1}^L m_l$. Taken together, the ADOs contain all non-Markovian and higher-order effects of the baths on the molecular dynamics. Each ADO is coupled to ADOs of higher and lower tier, $\rho^{\mathbf{n\pm1}_k,\mathbf{m}}$ and $\rho^{\mathbf{n},\mathbf{m\pm1}_l}$, where
\begin{equation}
\mathbf{n} \pm \mathbf{1}_k = [n_1, n_2 ,..., 1-n_k ,..., n_K],
\end{equation}
and
\begin{equation}
\mathbf{m} \pm \mathbf{1}_l = [m_1, m_2 ,..., m_l\pm 1 ,..., m_L].
\end{equation}
This culminates in a hierarchy of equations of motion that can be solved for the exact dynamics of the molecule. Note that in the HEOM formalism, the creation and annihilation operators are rewritten in a more convenient notation, such that $d^{+}_{i_k} = d^{\dag}_{i_k}$ and $d^{-}_{i_k} = d^{}_{i_k}$, and $\bar \sigma = -\sigma$. In our calculations, the basis set for each virtual phonon mode is limited to the 20 lowest energy states while the fermionic hierarchy is truncated after the $6^\text{th}$ tier, so that the observables are converged to the exact result.

The direct propagation of Eq.\,(\ref{HEOM}) is usually not computationally feasible when modeling scattering molecules due to the poor scaling of the dimensionality of the ADOs with respect to the number of vibrational basis states. An alternative is to reformulate the HEOM into a Schr\"odinger-like equation of motion for an extended wavefunction in twin space.\cite{yaling22} This then facilitates the use of a matrix product state (MPS) representation of the state of the system, \cite{shi18,borrelli19,takahashi24} which scales more favorably with respect to the basis size and eases the computational load. In this work, we employ an MPS representation of the state that is propagated in time via the time-dependent alternating minimal energy (tAMEn) propagator. \cite{dolgov19,borrelli21,taka24}.

\section{Hamiltonian Model Definition and Parametrization}\label{sec:Parametrization of Model Hamiltonian}

\subsection{Model Definition}
We choose a finite number of $N$ metal states and rewrite the electronic Hamiltonian, $\hat{H}_{\text{el}}^1$ ( \cref{eq:single-electron-hamiltonian}) in matrix form:
\begin{align}
    \hat{H}_{\text{el}}^1(x) =
\begin{pmatrix}
h(x) & V_{1}(x) & \cdots & V_{k}(x) & \cdots & V_{N}(x) \\
V_{1}(x) & \epsilon_1 & \cdots & 0 & \cdots & 0 \\
\vdots & \vdots & \ddots & \vdots& \vdots & \vdots \\
V_{k}(x) & 0 & \cdots & \epsilon_k & \cdots & 0 \\
\vdots & \vdots & \vdots & \vdots& \ddots & \vdots \\
V_{N}(x) & 0 & \cdots & 0 & \cdots & \epsilon_N \\
\end{pmatrix}. \label{eq:Hamiltonian-matrix-NA}
\end{align}
The parametrization of adsorbate energy curves takes inspiration from earlier work by one of the present authors on one-dimensional models for molecular junctions.\cite{erpenbeck2018, Erpenbeck2019, Erpenbeck2020} To model the nonadiabatic dynamics of hydrogen on Ge, we define neutral and negatively charged diabatic adsorbate energy landscapes, $U_0(x)$ and $U_1(x)$, respectively. In the neutral state, the interaction with the surface is described by a Morse potential, as, in the absence of charge transfer, the energy landscape of the neutral atom is dominated by Pauli repulsion effects, 
\begin{equation}
    U_0(x) =  D_e (\exp(-a(x-x_0)) - 1)^2 + c \label{eq:U0-hokseon}.
\end{equation}

In the charged state, we add a logistic function to describe the attractive interaction associated with charge transfer, such that 
\begin{equation}
    U_1(x) = U_0(x) + h(x) 
\end{equation}
where 
\begin{equation}
    h(x) =  \frac{D_1}{1 + \exp(-a' (bx - x_0'))} + c'. \label{eq:h-hokseon}
\end{equation}
When the adsorbate approaches the surface, the adsorbate state couples with the metal states via impurity-bath coupling terms:
\begin{equation}
    V_k(x) = \bar{a} \cdot \sqrt{w_k} \cdot A(x),
    \label{eq:A_k}
\end{equation}
where
\begin{equation}
    A(x) = \tilde{A} \left[ \dfrac{1-q}{2} \cdot \left(1 - \tanh\left(\frac{x-\tilde{x}_{0}}{L}\right)\right) + q\right] \label{eq:A-hokseon}
\end{equation}
describes the hybridization of the impurity in the wide band limit, $\sqrt{w_k}$ are bath discretization weights (with units of $\text{eV}^{1/2}$, \textit{vide infra}), and $\bar{a}$ is a geometry-independent rescaling factor (with units of $\text{eV}^{-1/2}$). The rescaling factor $\bar{a}$ depends only on the chosen electronic band width and thus serves as a convenient adjustment parameter when the model is parametrized for different band widths. This helps to avoid having to reparametrize $A(x)$ when the number of electronic states or the bandwidth is changed and facilitates the study of the convergence behaviour of the simulation results with respect to bandwidth and discretization shown in Fig.\,S3 of the Supporting Information (SI). The choice of modeling the dynamic charge transfer as an anionic resonance of hydrogen is based on the fact that the electron affinity level of hydrogen is more easily accessible than its ionization potential. It is further corroborated by Bader charge calculations (Fig.\,S5 in SI), which indicate that the hydrogen atom is partially negatively charged when adsorbed at the Ge surface.

\subsection{Bath Discretization for Semiconductors}

To render MQCD simulations of systems coupled to a continuum of states computationally feasible, the electronic bath needs to be discretized, for which many numerical techniques have been proposed.\cite{devegaHowDiscretizeQuantum2015} In previous work on MQCD on the NAH, the continuum of electronic states was efficiently discretized with a Gauss-Legendre method\cite{shenvi09, Gardner2023, gardner2023b, meng2024}. However, since Ge(111)$c(2\times8)$ is a semiconducting surface, it features a bandgap in its electronic band structure. Haldane and Anderson\cite{haldane1976} extended Anderson's original work\cite{Anderson1961} to materials with bandgaps. Inspired by this, we modify the Gauss-Legendre scheme of previous work\cite{Gardner2023, gardner2023b} such that the discretization includes an energy gap, $E_\text{gap}$, centered at the Fermi energy.  To achieve this, the energy of a discretized electronic state $k$, $\varepsilon_k$, in \cref{eq:Hamiltonian-matrix-NA}, is calculated via
\begin{equation}
\label{eq:GapGaussLegendre-states}
    \epsilon_k = \begin{cases}E_{\text{F}} - \left[\frac{\Delta E-E_{\text{gap}}}{2} (1+\epsilon_{\text L,N/2-k+1}) + E_{\text{gap}}\right]/2 &\text { if } k \leq N/2 \\
    E_{\text{F}} + \left[\frac{\Delta E-E_{\text{gap}}}{2} (1+\epsilon_{\text L,k-N/2}) + E_{\text{gap}}\right]/2&\text { otherwise }\end{cases}
\end{equation}
with conjugate weights
\begin{equation}
    w_k = \begin{cases}
        \Delta E w_{\text{L},N/2-k+1} /2 &\text { if } k \leq N/2 \\
        \Delta E w_{\text{L},k-N/2} /2 &\text { otherwise }
    \end{cases}.
\end{equation}
Here, $\epsilon_{\text{L},i}$ and $w_{\text{L},i}$ represent the knot points and weights, respectively, derived from Legendre quadrature over the interval $[-1,1]$, using $N/2$ knots. The parameter $E_{\text{F}}$ indicates the Fermi level, usually set to 0 eV or the centre of the continuum. The total number of metal states, $N$, must be an even number for this formulation. Simply put, the Gauss-Legendre discretization is done independently for the valence and conduction bands, such that a gap is formed between the two sets of knots that constitute the bands. Other discretization schemes are possible and modification of those schemes by insertion of a bandgap would be equally straightforward. For example, we can discretize the Hamiltonian with a simple trapezoidal scheme, which allows us to assess the impact of the gap-inserting discretization scheme on the resulting density of states. Upon comparison of the resulting density of states obtained with the two schemes (Fig.\,S2 in SI), we find no substantial variation between the two schemes.

\subsection{Model Parameterization}

\begin{figure}[h]
    \centering
    \includegraphics[width=3in]{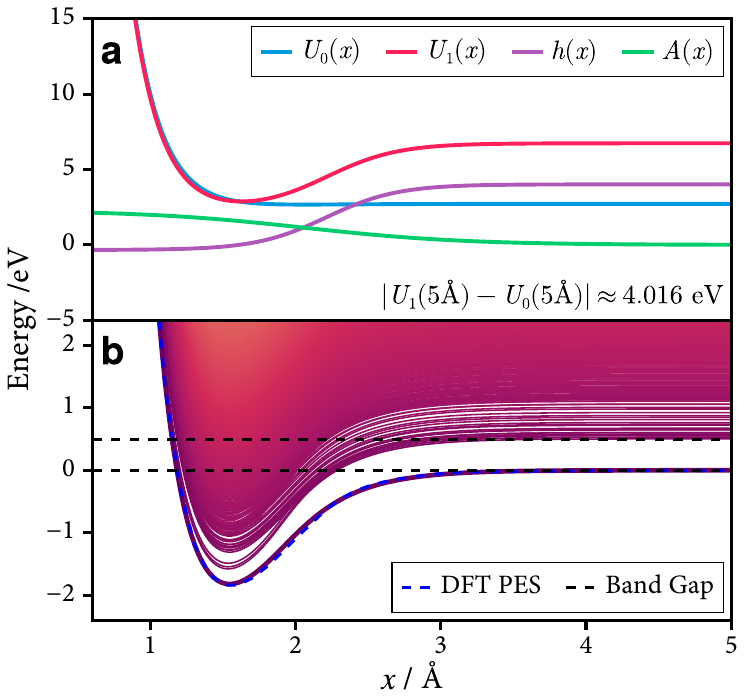}
    \caption{Panel \textbf{a}: The diabatic adsorbate states and coupling strength given by \cref{eq:U0-hokseon}-\cref{eq:A-hokseon} as a function of hydrogen atom height $x$ with corresponding conjugate parameters given in \cref{tab:Hokseon-Parameters}. Panel \textbf{b}: The adiabatic potential energy spectrum for the 2000 lowest energy many-electron states. The spectrum is produced with the gap-inserting Gauss-Legendre method using 150 states, 75 electrons and a bandwidth of $\Delta E =$ 50\,eV. The inserted bandgap has width $E_{\text{gap}} = 0.49$\,eV. The Fermi level is $E_\text{F}= 0 $\,eV. The rescaling parameter for the couplings is  $\bar{a} = 0.479\,\text{eV}^{-1/2}$.}
    \label{fig:diabatic-adiabatic-PES}
\end{figure}

To specifically describe the interaction of a hydrogen atom with a Ge(111)$c(2\times8)$ surface, we impose  the following constraints to the Hamiltonian and the state-independent potential:
\begin{itemize}
\item The total ground state energy landscape is fitted to match the hydrogen-germanium interaction energies obtained with density functional theory (DFT):
\begin{equation}
    E_{\mathrm{DFT}}(x) \stackrel{!}{=}
 U_0(x) + \sum_{k=1}^{N/2} \lambda_k(x)
\end{equation} 
\item Second, in the interaction free region, the energy difference between the diabatic curve for the charged and neutral hydrogen atom, $h(x\rightarrow\infty)=U_1 - U_0$, needs to be close to the difference between the work function of Ge, which is 4.78\,eV\cite{dillon1957} and the electron affinity of atomic hydrogen (0.75\,eV).\cite{Haffinity}

\item Third, the NAH model is parametrized such that the gap between the adiabatic ground state and the first excited state closes at a point closer to the surface than the adsorption minimum. We set $h(x)= -E_{\mathrm{gap}}/2$, where $x$ corresponds to an adsorption height of $x \approx 1.15$\,\AA. At this height, the potential energy of the DFT-calculated adiabatic ground state is exactly the experimentally measured Ge(111)$c(2\times8)$ bandgap of 0.49\,eV\cite{feenstra2006} above the dissociated limit, i.e., $E_{\mathrm{groundstate}}(x=1.15)=E_{\mathrm{DFT}}(\infty)+E_{\mathrm{gap}}$, and a particle with sufficient kinetic energy to overcome the gap can reach the region where the adsorbate state $h(x)$ becomes degenerate with the upper edge of the valence band. This is depicted in Fig. S4 (a zoomed-in version of \cref{fig:diabatic-adiabatic-PES}b in the SI. When particles reach $x\approx1.15$\,\AA, excitations between the many-electron ground state and excited states can readily occur.

\end{itemize}
The fitting of model parameters is performed with the package \texttt{BlackBoxOptim.jl}\cite{vaibhav_kumar_dixit_2023_7738525} (version 0.6.3). The final parametrized adsorbate functions are visualized in \cref{fig:diabatic-adiabatic-PES}(a), and the individual parameters of $U_0(x)$, $U_1(x)$, and $A(x)$ are given in \cref{tab:Hokseon-Parameters}.

\begin{table}[ht]
   \caption{Parameters for the adsorbate model defined in eqs. \cref{eq:U0-hokseon}-\cref{eq:A-hokseon}.}
   \label{tab:Hokseon-Parameters}
   \centering
   \renewcommand{\arraystretch}{1.5} 
   \setlength{\tabcolsep}{10pt} 
   \begin{tabular}{ccc} 
   $U_0(x)$ & $h(x)$ & $A(x)$\\
   \toprule
   $D_e \quad 0.0502 \text{ eV}$ & $D_1 \quad 4.351 \text{ eV}$ & $\tilde{A} \quad 2.28499 \text{ eV}$  \\
   $a \quad 2.39727 \text{ \AA}^{-1}$ & $a' \quad 3.9796 \text{ \AA}^{-1}$ & $L \quad 1.1012 \text{ \AA}$ \\
   
   $x_0 \quad 2.0727 \text{ \AA}$ & $x_0' \quad 2.2798 \text{ \AA}$ & $\tilde{x}_0 \quad 2.0589 \text{ \AA}$ \\

   $c \quad 2.67532 \text{ eV}$ & $c' \quad -0.33513 \text{ eV}$ & $q \quad 2.26\times 10^{-16}$ \\
                               & $b \quad 1.02971$           &       \\
   \bottomrule
   \end{tabular}
\end{table}

Given a discretized bath with a bandgap and the above defined adsorbate diabatic potentials, \cref{fig:diabatic-adiabatic-PES}(b) shows the  2000 lowest energy states produced by swapping occupied with unoccupied states compared to the ground-state Slater determinant. At large distances, the ground state curve and the first excited state show an energy difference equal to the experimentally determined bandgap value of 0.49\,eV\cite{feenstra2006} as was enforced during parametrization. At the equilibrium height, the ground state is still energetically well separated from the excited state. In the repulsive region{---}at energies close to the bandgap{---}the energy difference between the adiabatic ground state and the first four excited states becomes very small as it was proposed in previous studies.\cite{kruger2024b, zhuMechanisticInsightsNonadiabatic2024} For an electronic bath with an effective (combined valence and conduction) bandwidth of 49.51\,eV, the rescaling factor $\bar{a}$ in \cref{eq:A_k}, used for \cref{fig:diabatic-adiabatic-PES}(b), is set to 2.5\,a.u. ($\approx 0.479\,\text{eV}^{-1/2}$). We observe that the parametrization remains robust across a range of constant electronic effective bandwidths, with appropriate tuning of $\bar{a}$. Specifically, larger bandwidths require smaller values of $\bar{a}$, while narrower bandwidths require larger rescaling factors. This behavior is illustrated in Fig. S1 of the SI.

\section{Computational Details}

\subsection{Density Functional Theory reference data}
The density functional theory (DFT) input data for the 1D energy landscape were generated with the electronic structure code FHI-aims.\cite{blum_ab_2009} 
We used the standard "tight" default basis set for Ge and light basis set for hydrogen. Convergence tests with tight basis on hydrogen showed no significant change. The Brillouin-zone was sampled with a $12\times4\times1$  Monkhorst-Pack $\bm k$-point grid.\cite{monkhorst76} 
Relativistic corrections were incorporated with the ZORA scheme.\cite{ZORA} 
We used a Gaussian occupation smearing width of 0.1 eV and the PBE exchange-correlation functional.\cite{pbe1,pbe2} Convergence criteria for the energy, eigenvalues, and density were $10^{-6}$\,eV, $10^{-2}$\,eV, and $10^{-5}$\,e/$\text{a}_0^3$, respectively. To account for the open-shell nature of the hydrogen atom, we performed spin-polarized DFT calculations. The Ge(111)$c(2\times8)$ surface was modeled by a slab with eight layers and four additional Ge atoms added onto the top-layer as described in Ref.~\citenum{kruger2023}. The resulting structure is depicted in \cref{fig:HGe_structure}. A vacuum slab of  40\,{\AA} and a dipole correction were used. The Ge atoms constituting the lowest layer in the slab were saturated with hydrogen atoms. To obtain a static grid of interaction energies, we placed the hydrogen atom 5\,{\AA} above the rest-atom and then varied the $z$-coordinate in 0.1\,{\AA} increments until a vertical distance of 0.8\,{\AA} was reached. Thus, the input data set consists of 52 data points.

We also conducted a charge analysis using the Quantum Theory of Atom In Molecule (AIM) theory proposed by Bader.\cite{bader} For this purpose, the AIM utility was employed to analyze the charge density obtained from additional DFT calculations performed with the ABINIT package.\cite{Gonze2020}
A slab model was built, consisting of a $2\times2$ germanium surface cell, comprising eight Ge layers symmetrically arranged along the z-axis, with a vacuum region 18 {\AA} thick. A hydrogen atom was placed above the surface at varying distances from the outermost germanium atoms. Similarly to the FHI-aims case, these calculations used spin-polarized wavefunctions and the PBE functional. Furthermore, norm-conserving pseudopotentials, a plane-wave energy cutoff of 40 Ha, and a Gaussian occupation smearing of 0.01 eV were employed.

\subsection{Dynamics Simulations}
The MQCD calculations were performed with the \texttt{NQCDynamics.jl}\cite{gardner2022, gardner2023b} package (version 0.13.4) and its companion package, \texttt{NQCModels.jl} (version 0.8.19) , written in the Julia programming language. At the beginning of all simulations, the hydrogen atom was placed in the interaction free region of the potential at a height of 5\,{\AA} above the surface. The initial momentum vector of the hydrogen atom always points towards the surface and its magnitude is determined by a preset kinetic energy which is in line with previous scattering simulations for hydrogen atom scattering from insulators,\cite{hertl21b} metal,\cite{hertl21a, hertl22b} and semiconductor surfaces.\cite{kruger2023, kruger2024} The electronic population was sampled from a Fermi-Dirac distribution with a temperature of $300$\,K. For the discretization, we used the gapped Gauss-Legendre discretization detailed in \ref{sec:Parametrization of Model Hamiltonian}A. Numerical integration of the equations of motion was performed with a time-step $\Delta t = 0.05$\,fs. The trajectories were terminated if either the hydrogen reached a final height larger than 5\,{\AA} or the simulation time exceeded 1001\,fs. The band of the electronic states was represented by 150 electronic states, and the width of the band was set to 50\,eV for which we found that convergence was achieved (see Section S3 in the SI for details). 

To study the influence of the initial kinetic energy on the mixed quantum classical dynamics simulations, we explored a wide range of initial kinetic energies ranging from 0.1\,eV to 7\,eV. Two MQCD techniques were used:  Ehrenfest Dynamics and IESH.\cite{shenvi09} The specific implementation details for the propagation algorithm of Ehrenfest dynamics and IESH can be found in Ref.\citenum{gardner2022} and Ref.\citenum{Gardner2023}, respectively. 
Only a single Ehrenfest dynamics trajectory was simulated for each initial condition. Given the stochastic nature of IESH, however, we ran 500 trajectories in order to acquire mean final kinetic energies of scattered hydrogen atoms. The statistical uncertainty was quantified via a 95\% confidence interval. To achieve statistical convergence for the energy loss distributions and associated sticking events, we ran the following numbers of trajectories: 509,680 for \cref{fig:Exp-IESH-0.37}, and 1,065,167, 1,071,077, and 1,357,661 for (\textbf{a}), (\textbf{b}) and (\textbf{c}) in \cref{fig:Exp-IESH}, respectively. Furthermore, to study the energy dependence of sticking with IESH, we launched 87,000 trajectories at several kinetic energies above and below the bandgap (Figure S4 in the SI). To compare the nonadiabatic behavior between IESH and HEOM for the proposed H/Ge model, we launched trajectories with initial position and momenta sampled from a Wigner distribution.

\section{Results and Discussion}

\subsection{Comparison between IESH and Ehrenfest dynamics}

\begin{figure}[H]
    \centering
    \includegraphics[width=3.3in]{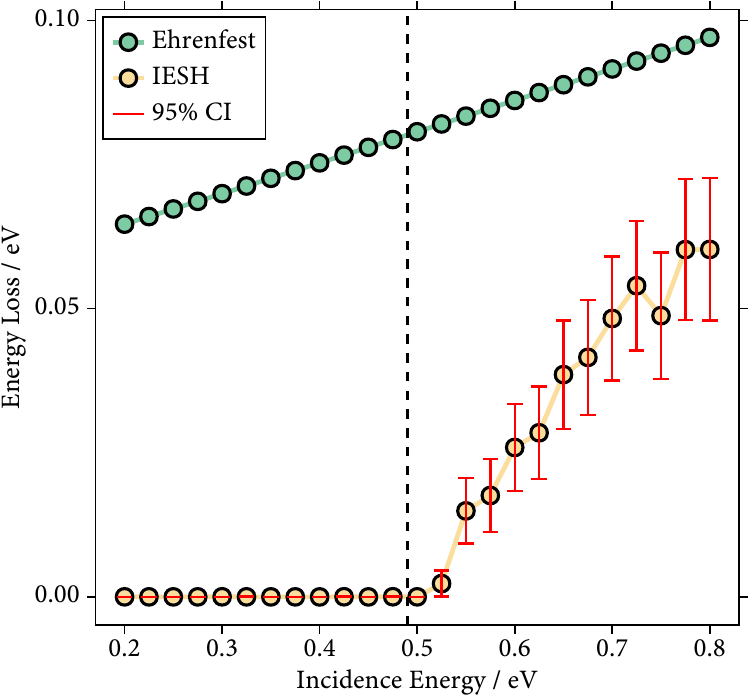}
    \caption{ Average kinetic energy loss of scattered hydrogen atoms obtained via IESH and Ehrenfest simulations with incidence energies $E_i$ ranging from 0.2 to 0.8\,eV. Each IESH data point is averaged over 1000 trajectories (60,000 for $E_i = 0.5$ and $0.525$\,eV), while the deterministic Ehrenfest method uses a single trajectory. The blue dashed line represents zero energy loss (elastic scattering), black solid line references an absolute constant energy loss, and the vertical black dashed line indicates the bandgap, $E_{\text{gap}} = 0.49$\,eV.}
    \label{fig:IESH-Ehrenfest-final-kinetic-loss}
\end{figure}

With the intent to assess which MQCD method is more suitable for describing the experimentally observed nonadiabatic electron transfer into the conduction band of Ge, we performed hydrogen atom scattering simulations incorporating nonadiabatic effects with IESH and Ehrenfest. The mean kinetic energy loss as a function of initial kinetic energy is reported in \cref{fig:IESH-Ehrenfest-final-kinetic-loss}.

The IESH simulations only yield kinetic energy loss upon scattering when the initial kinetic energy of the impinging hydrogen atoms exceeds the bandgap, i.e., ($E_i > E_{\text{gap}} = 0.49$\,eV). When $E_i < E_{\text{gap}}$, the hydrogen atoms scatter elastically. This can be explained with the adiabatic potential energy spectrum shown in \cref{fig:diabatic-adiabatic-PES}(b). The energy gap between the adiabatic ground and first excited states closes at a potential energy value equal to the bandgap, i.e., 0.49\,eV. Thus, the closer the hydrogen atom approaches this region in the PES, the more likely it is that surface hops occur. At small kinetic energies, the nonadiabatic energy dissipation channel is closed as the incoming hydrogen atoms will not reach the repulsive region where the adiabatic ground state and the first excited state intersect and therefore the hopping probabilities are small. 

\begin{figure}[H]
    \centering
    \includegraphics[width=3.3in]{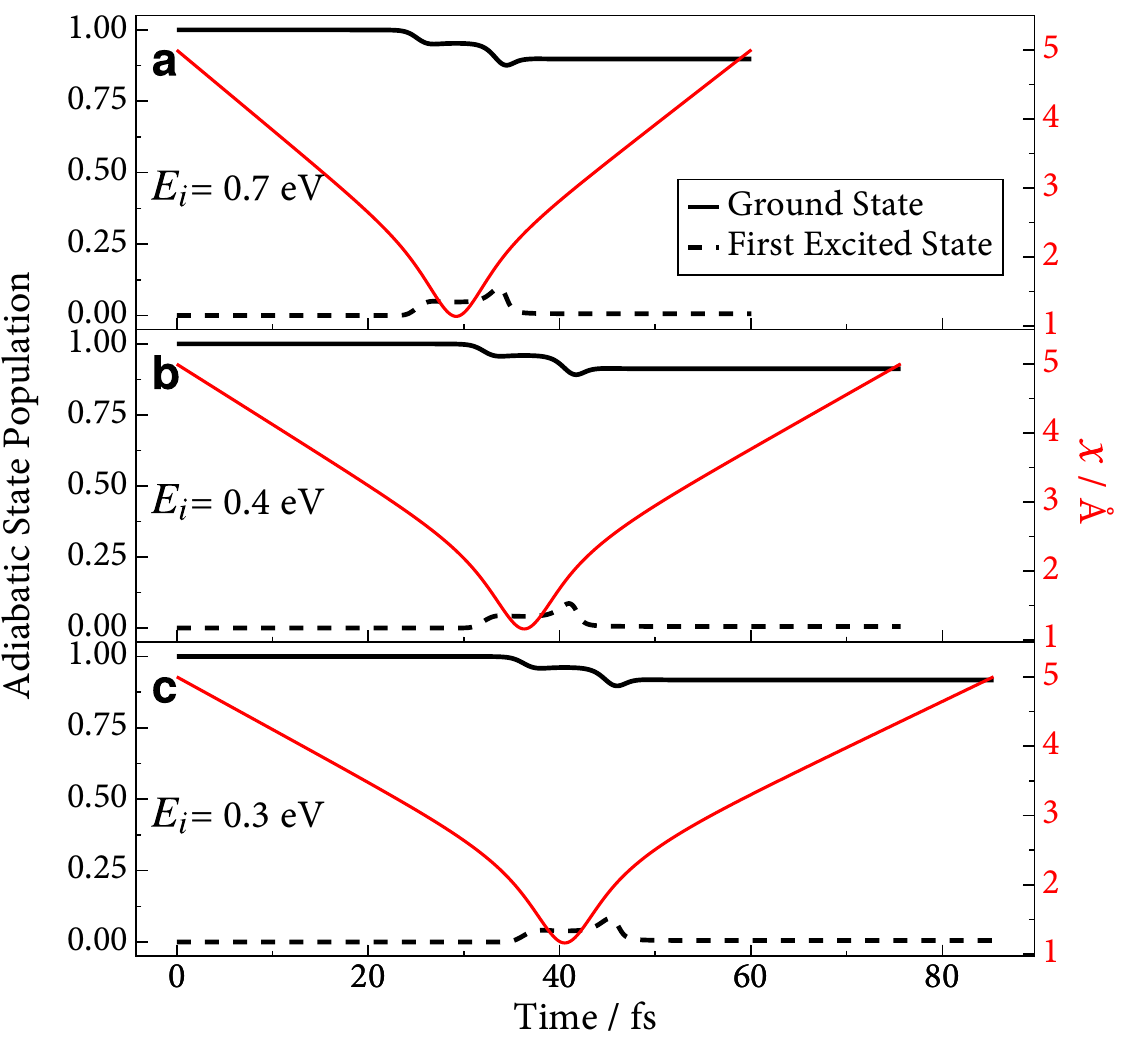}
    \caption{Ehrenfest adiabatic electronic state population and height of the hydrogen atom during scattering at an initial height of $x_0=$5\,\AA, with three initial kinetic energies $E_i=$ 0.7\,eV(\textbf{a}), 0.4\,eV(\textbf{b}) and 0.3\,eV(\textbf{c}). The black curves are the adiabatic population and the red curve describes the hydrogen atom position (with respect to the right ordinate).}
    \label{fig:Ehrenfest-position-adiabatic-state-population}
\end{figure}

In contrast, ED simulations yield nonadiabatic energy loss at any initial kinetic energy across the range of incidence energies from 0.2 to 0.8\,eV. This contradicts experimental observations.\cite{kruger2023}

To understand the origin of linear increase of energy loss across the range of investigated energies in \cref{fig:IESH-Ehrenfest-final-kinetic-loss}, we investigated the populations of the adiabatic ground state and first excited state across the scattering trajectory for three different kinetic energies. At the beginning of each simulation, only the adiabatic ground state is populated (\cref{fig:Ehrenfest-position-adiabatic-state-population}). The analysis indicates that the first excited state always starts to contribute to the effective electronic potential of Ehrenfest dynamics once the hydrogen atom has reached a height of approximately 2.5\,\AA\,from the surface. Upon further approach, the contribution of the first excited state to the effective potential increases up to a maximum of $\sim$8\% until it starts to reduce once the hydrogen atom recoils from the surface. This observation holds irrespective of whether or not the incidence energy is above or below the bandgap. Note, however, that the contribution of the adiabatic ground state monotonically decreases over the entire course of the trajectory and ends up at a population of about 88\%. Hence, excited electrons are created in the conduction band during the scattering, which immediately affect the mean-field potential that steers the hydrogen atom dynamics. Given that the final contribution of the first excited state is smaller than 12\% we can assert that not only the first excited state but also higher excited states contribute to the mean field potential. This mixing effect is slightly stronger for higher initial kinetic energies because the hydrogen atom manages to come closer to the surface but the overall picture stays the same through the entire investigated kinetic energy range. 

We conclude that ED simulations are not able to describe the bandgap threshold in collision-induced electron excitations into the conduction band in the H/Ge model,  whereas IESH correctly captures this effect. In the subsequent sections, we will therefore only focus on IESH simulation results.

\subsection{Low incidence energy regime}

\begin{figure}[H]
    \centering
    \includegraphics[width=3.3in]{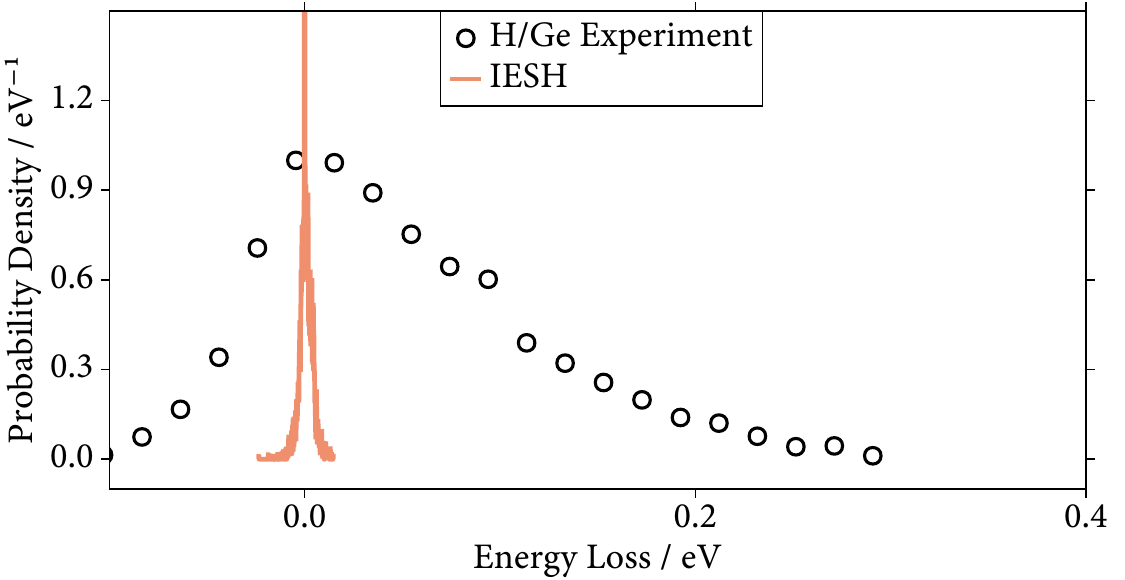}
    \caption{Probability density of scattered kinetic energy loss simulated by IESH with incidence energies $E_i = 0.37$~eV at temperature $T = 300$~K, compared with experimental data for hydrogen atom scattering from Ge(111)$c(2\times8)$.\cite{kruger2023} The IESH distribution is constructed using a Gaussian kernel density estimate (KDE) with standard deviation $\delta_{\text{sd}} = 0.00005$.}

    \label{fig:Exp-IESH-0.37}
\end{figure}

Complementary to the mean kinetic energy loss for scattered hydrogen atoms depicted in \cref{fig:Ehrenfest-position-adiabatic-state-population}, we also report kinetic energy loss distributions which can be measured experimentally.\cite{buenermann18}  The IESH simulations yield an energy loss distribution in the form of a sharp peak centered precisely at zero energy loss (\cref{fig:Exp-IESH-0.37}). This means that at an incidence energy of 0.37 eV, only elastic scattering occurs. This observation aligns with the general trend shown in \cref{fig:IESH-Ehrenfest-final-kinetic-loss}, which demonstrates that the IESH simulation predicts no significant energy loss until the incidence energy exceeds the gap energy ($E_i > E_{\text{gap}}$). The differences between simulation and experiment are expected because it was shown that the experimental peak below the bandgap is caused by excitation of lattice vibrations,\cite{kruger2024b} a physical process which our simple model does not capture. In other words, the IESH simulations are consistent with expectations; this MQCD method does not predict any nonadiabatic energy loss in the kinetic energy distribution for energies below the bandgap.

\subsection{High incidence energy regime}

\begin{figure}[H]
    \centering
    \includegraphics[width=3.3in]{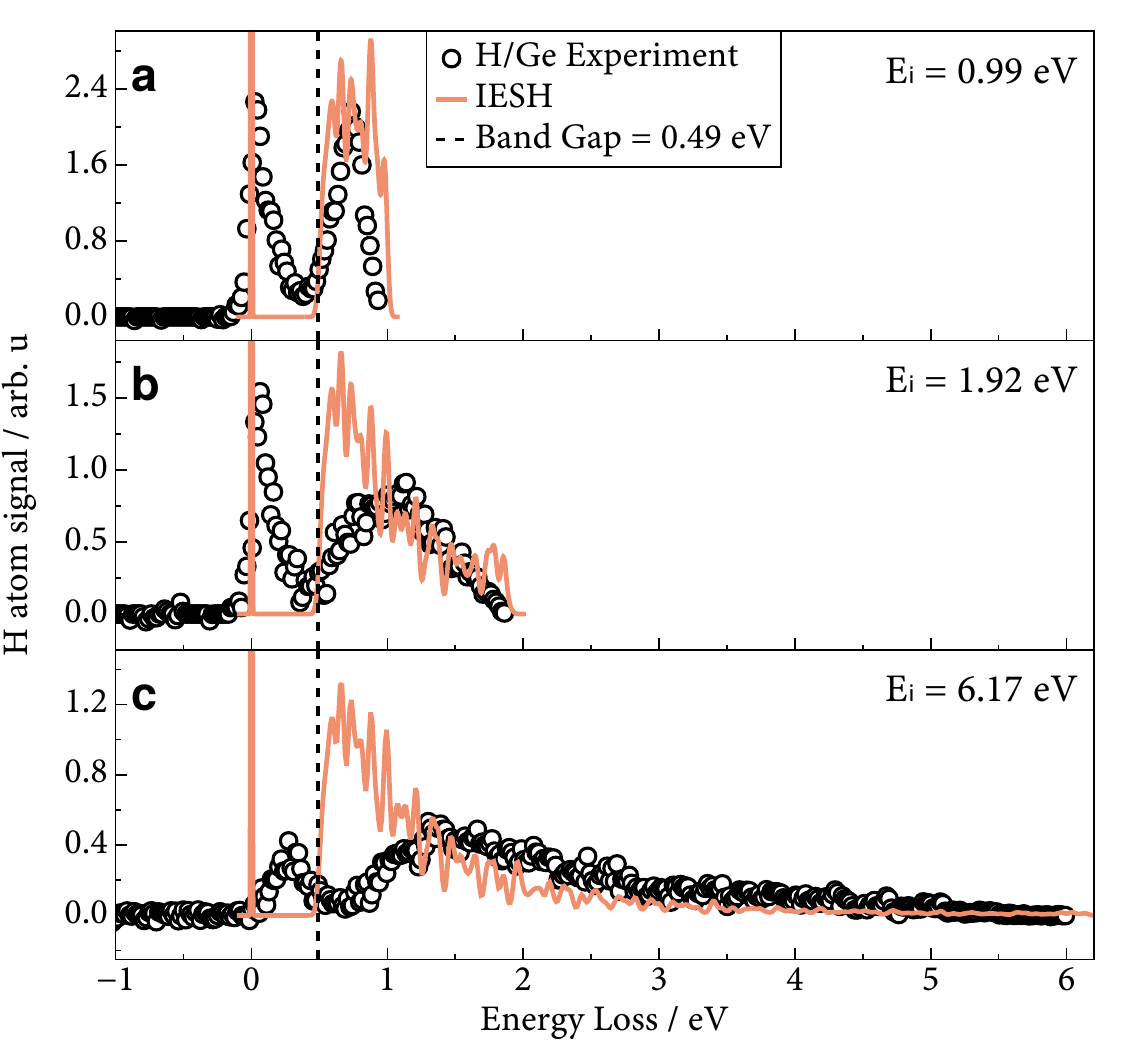}
    \caption{Probability density of kinetic energy loss simulated by IESH with incidence energies $E_i = 0.99$~eV(\textbf{a}), 1.92~eV(\textbf{b}) and 6.17~eV(\textbf{c}) at temperature $T = 300$~K, compared with experimental data for hydrogen atom scattering from Ge(111)$c(2\times8)$. The inelastic and elastic distributions from IESH are independently constructed via a Gaussian kernel density estimate (KDE) with standard deviations $\delta_{\text{sd}} = 0.02$ and $0.003$, respectively. The experimental data is normalised with respect to the area under the curve. The fraction of inelastic IESH scattering events is $11.5\%$(\textbf{a}), $16.88\%$(\textbf{b}) and $26.46\%$(\textbf{c}).}
    \label{fig:Exp-IESH}
\end{figure}

We now turn to initial kinetic energies above the bandgap, where experimental inelastic kinetic energy loss distributions for hydrogen atom scattering from Ge(111)$c(2\times8)$ have a bimodal shape.\cite{kruger2023, kruger2024, kruger2024b} While the peak at energy losses smaller than the value of the bandgap of Ge(111)$c(2\times8)$ has been well-characterized and assigned to lattice excitations, it was proposed that the other peak in the energy loss regime originates from electronic excitations into the conduction band. 
We therefore conduct IESH simulations at three different experimentally accessible initial kinetic energies ($E_i = 0.99$, 1.92, and 6.17~eV), and the results are shown in \cref{fig:Exp-IESH}.  As anticipated, IESH simulations predict an inelastic peak located at energies above the gap energy $E_{\text{gap}}$ and also a narrow peak centered at 0\,eV demonstrating elastic scatterings for each $E_i$. The elastic component is artificially narrow due to the omission of phonons, which contribute to the quasi-elastic peak seen in the experiments. For all applied kinetic energies that exceed the bandgap value, we find that the ratio between the inelastic scattering and elastic scattering events is lower in comparison to the experiment. The simulated ratios are 0.11, 0.17, and 0.26 for 0.99\,eV, 1.92\,eV, and 6.17\,eV, respectively. The experimental ratios reported by Kr\"uger \textit{et al.} are 1.2, 2.2, and 10 for 0.99\,eV, 1.92\,eV, and 6.17\,eV, respectively.\cite{kruger2023} We assign this difference to the simplicity of the model and the single wide band limit representation of the electronic structure, which we will discuss in detail below. We will first focus on the discussion of the shape of the inelastic peak.

A comparison with the experimental data shows that the simulation with lower incidence energy provides good agreement for the nonadiabatic component. As shown in \cref{fig:Exp-IESH}(\textbf{a}), one can see that despite the simplicity of the underlying model the IESH inelastic simulation already manages to reproduce the experimental width and height of the probability density.

In contrast, for the higher kinetic energy values $E_i = 1.92$\,eV and $E_i = 6.17$\,eV, shown in \cref{fig:Exp-IESH}(\textbf{b}) and \cref{fig:Exp-IESH}(\textbf{c}), respectively, more pronounced differences start to emerge. For $E_i$=1.92\,eV, the maximum of the IESH-based kinetic energy loss distribution sits at an energy loss close to 0.49\,eV, i.e., the bandgap of Ge(111)$c(2\times8)$, whereas the maximum of the experimental peak is located around 1\,eV. Moreover, the high-energy part of the experimental distribution is symmetric around its maximum, whereas the simulated kinetic energy distribution comprises a sharp increase in intensity to its maximum at an energy loss value close to 0.49\,eV, i.e., the bandgap of Ge(111)$c(2\times8)$ followed by a decaying tail towards higher energy loss values.  For $E_i=6.17$\,eV, the observed differences in the shape of the experimental and computational distribution are very similar to the differences for the $E_i=1.92$\,eV curves but more exaggerated: Here, the experimental distribution is broadly spread, mostly covering energy losses from 1\,eV to 4\,eV. However, the IESH simulations in the one-dimensional model deviate significantly, with most trajectories ending up with an energy loss within a narrow range from 0.49\,eV to 1\,eV and thus showing a much higher probability in this energy loss range compared to experiments.

It is worth highlighting that we do not aim for a quantitative modeling of the scattering experiments but rather for an assessment of the different MQCD methods proposed to be suitable for the description of strong nonadiabatic effects. There are several reasons why the one-dimensional model cannot provide quantitative predictions of experimental kinetic energy loss distributions: First, we restrict our model to one degree of freedom that describes normal incidence scattering at the Ge rest atom of the surface. The scattering geometry in our model corresponds to a central impact of one hydrogen atom on a single Ge atom with normal incidence angle, with the exit channel being normal incidence as well. In the experiment, however, this particular scattering geometry cannot be detected. Instead, commonly, specular scattering for an angle $45^\circ$ with respect to the surface normal is recorded. This difference between our simulations and the experimental conditions prevents any quantitative comparisons with experiment. Due to the reduction to one degree-of-freedom our model also does not capture the distortion of the surface geometry caused by the impact of the hydrogen atom and the concomitant charge redistributions and changes in the electronic structure of the germanium surface. 
Secondly, in our model, the probability for colliding with a surface atom is always one, whereas this is not the case in reality. In fact, the likelihood for an incoming hydrogen atom to penetrate the first surface layer and explore deeper surface areas increases with decreasing packing density.\cite{hertl22b} The reconstructed Ge(111)$c(2\times8)$ surface is relatively open and therefore subsurface dynamics cannot be ruled out{---}in particular at increasing initial kinetic energies, which may be one of the reasons why the $0.99$\,eV simulations agree best with experiment. Another simplification is the treatment of the band structure. We assume a constant density of states with constant coupling to the adsorbate through the entire spectral range except for the band gap of the Ge(111)$c(2\times8)$ surface. The model neglects variations in coupling due to varying contributions of s and p bands at different energies. In conjunction with the previous argument on subsurface scattering, when H atoms with high kinetic energy such as 1.92\,eV or 6.17\,eV explore the deeper region of the substrate, excitations of electrons across the bandgap of the germanium bulk, 0.66\,eV\cite{kittel}, cannot be ruled out. Moreover, our treatment of the electronic states also neglects the contribution of interband excitations at higher energies. IESH neglects electron-electron scattering and excited electrons have longer lifetimes han expected. Therefore, there is a finite probability that excited electrons might deexcite again during the same scattering event to transfer energy back to the hydrogen atom, leading to a scattering event with two nonadiabatic transitions but almost no energy loss. Additionally, the probability for a single collision is increased if the colliding particle impacts a surface atom centrally and therefore, we miss out on a major fraction of the scattering dynamics. Moreover, in our simulations, the hydrogen atom cannot dissipate kinetic energy via lattice excitations, which explains why the maxima of our kinetic energy distributions are located close to the gap energy and do not shift to slightly higher energy losses with higher initial kinetic energies. Energy dissipation to lattice vibrations may broaden the final kinetic energy distribution as the impinging adsorbates probe a statistical distribution of displaced lattice geometries. However, it was previously shown that phonons do not strongly affect energy loss distributions for H atom scattering from metal surfaces.\cite{grimmelmann80, hertl21a, martin-barrios2022} We will discuss this further in the next subsection. Finally, the only first-principles component which enters our model is the DFT ground state energy but all other ingredients are phenomenological. Constrained density functional theory\cite{meng2022, meng2024} for instance could provide \textit{ab-initio} diabatic states and couplings from which one could construct Haldane-Anderson type Hamiltonians for MQCD simulations for hydrogen atom scattering from Ge(111)$c(2\times8)$. In consequence, we assert that despite the simplicity of our model Hamiltonian which we used for MQCD simulations, the final results are already respectable and correctly capture the key physical phenomenon that leads to nonadiabatic energy dissipation.

\subsection{Comparison to HEOM}
\label{sec: Comparison to HEOM}

In this section, the IESH scattering data is validated against numerically exact data computed via HEOM. The quantum wavepacket in the HEOM simulations is initiated as a Gaussian wavepacket with an average kinetic energy $p_\text{ini}^2/{2m}$ towards the surface,
\begin{equation}
\Psi(x) = \frac{1}{(2\pi \sigma^2)^\frac{1}{4}}\exp\big[-\frac{(x-x_\text{ini})^2}{4\sigma^2}+i\frac{p_\text{ini}}{\hbar}(x-x_\text{ini})\Big],
\label{eq: HEOM wavepacket}
\end{equation}
where $\sigma$ determines the width of the wavepacket, and $x_\text{ini}$ and $p_\text{ini}$ are the initial center position and momentum, respectively. The vibrational basis set is given by a sine-DVR basis on a grid that ranges from $x_0 = 0$\AA\ to $x_{N_\text{DVR}} = 10$\AA, with $N_\text{DVR} = 315$ basis states. Unless otherwise specified, we take $x_\text{ini} = 6$\AA. To facilitate a direct comparison between the methods, the initial conditions of each IESH trajectory are now sampled from the corresponding Wigner distribution for the Gaussian wavepacket,
\begin{equation}
\Phi(x,p) = \frac{1}{\hbar \pi}
\
\exp\Big[-\frac{(x-x_{ini})^2}{2\sigma^2}\Big]
\
\exp\Big[- \frac{2\sigma^2}{\hbar^2} (p - \hbar p_{ini})^2\Big].
\label{eq: Wigner distribution}
\end{equation}
A direct application of the HEOM approach is limited to initial energy distributions for the wavepacket that are much wider than the narrow distributions achieved in hyperthermal atom scattering experiments. Since the kinetic energy loss distribution is strongly dependent on the initial wavepacket width, a more meaningful comparison in this section is obtained by benchmarking the final kinetic energy distribution directly. In the future, a more refined treatment could employ, for example, a flux correlation analysis. \cite{Gatti23}

\begin{figure}[H]
    \centering
    \includegraphics[width=3.3in]{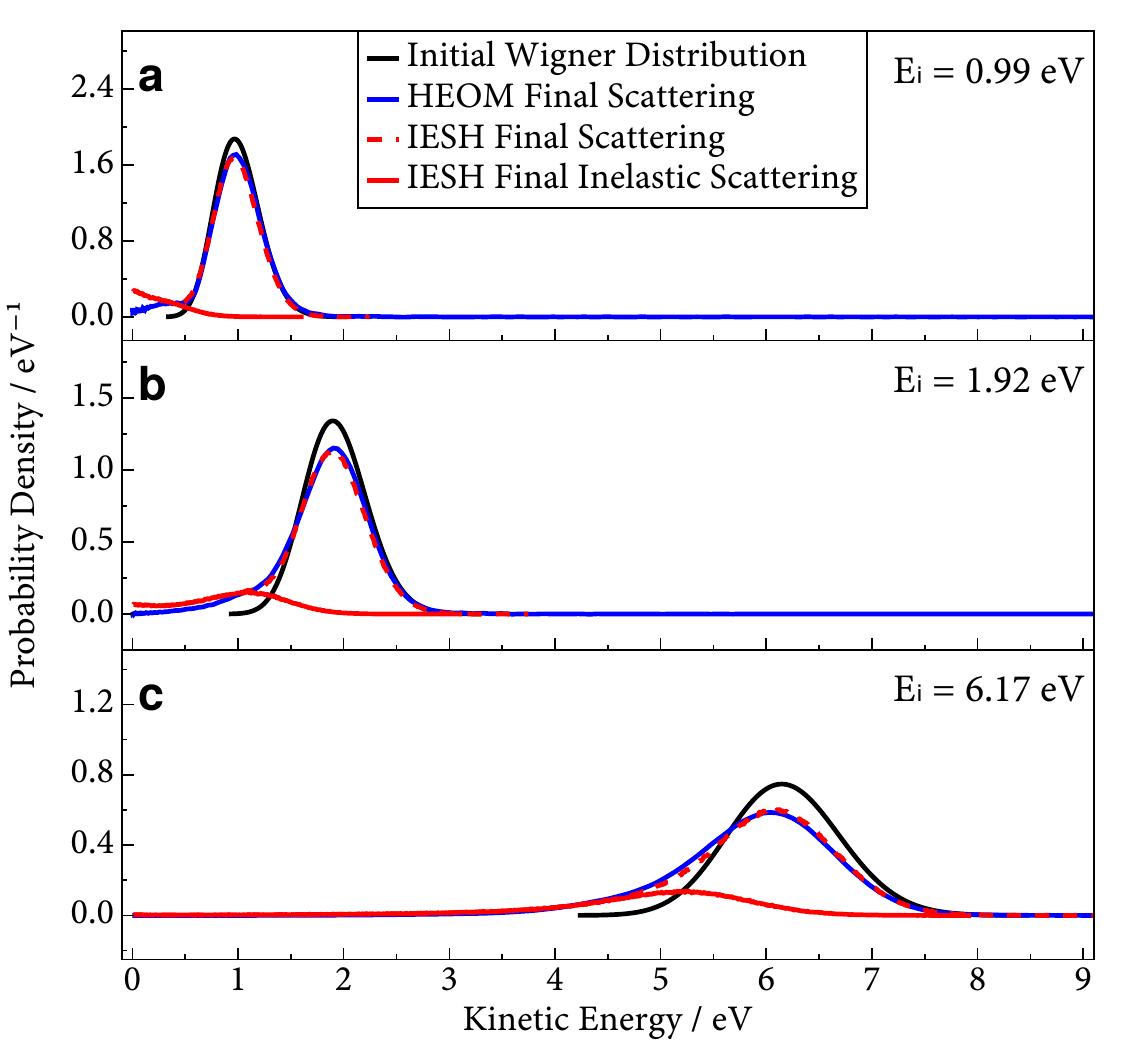}
    \caption{Probability density of the final kinetic energy for hydrogen atom scattered from Ge(111)$c(2\times8)$, simulated using IESH and HEOM. Results are shown for incidence energies $E_i = 0.99$~eV(\textbf{a}), 1.92~eV(\textbf{b}), and 6.17~eV(\textbf{c}) at $T = 300$~K. The initial states are the Wigner distribution with width $\sigma = 0.4$\,a.u in real space (black line). The full scattering distribution (red dashed line), which includes both elastic and inelastic contributions, and the inelastic scattering distribution (red baseline) from IESH were constructed using a Gaussian kernel density estimate (KDE) with a standard deviation $\delta_{\rm sd} = 0.005$. The probabilities for inelastic scattering events are 0.112, 0.170, and 0.264 for 0.99~eV, 1.92~eV, and 6.17~eV, respectively.
    }
    \label{fig:IESH-HEOM}
\end{figure}

A comparison between the final kinetic energy distributions calculated via IESH and HEOM is shown in Fig.\,\ref{fig:IESH-HEOM} for three different initial kinetic energies of the atom. If all scattering events were purely elastic, the simulated distributions would coincide exactly with the initial Wigner distributions for every incidence kinetic energy. The observed deviations from the initial Wigner distributions arise from a small fraction of inelastic scattering processes. We find a good agreement between IESH and HEOM for a variety of different initial wavepacket widths $\sigma$ which we show in Figs. S7 \& S8 in the SI. The close agreement between the final scattering results obtained from the IESH and HEOM methods indicates that both approaches capture similar elastic and inelastic behavior. It further demonstrates the validity of IESH as an approximate non-adiabatic simulation technique for energetic transition over energy ranges of the scale of several hundreds of meVs such as band gaps. The absence of the double peak structure in Fig.\,\ref{fig:IESH-HEOM} in the final kinetic energy distribution is a result of the narrow spatial width $\sigma$ of the Gaussian wavepacket. As one can infer from \cref{eq: HEOM wavepacket,eq: Wigner distribution}, a spatially narrow wavepacket results in a broad distribution in momentum space. When we increase $\sigma$ a bimodal structure is found, as is shown in Fig.\,S9 in the SI.  A broader wavepacket in real space makes the HEOM calculations computationally more demanding and unfeasible if we attempt to simulate energetic widths of the H atom pulses in the experiment, which are on the order of meV.\cite{buenermann18}

We also performed HEOM calculations with the inclusion of a phononic bath based on the approach proposed by Ke \textit{et al.}.\cite{ke2022} We find a weak influence of the phonons on the overall final kinetic energy distribution for a variety of coupling strengths. This is in accordance with previous simulations of H atom scattering off a W(110) surface, where the impact of phonons{---}either modelled explicitly via full-dimensional potentials or in the form of a phononic bath{---}on the overall shape of the energy loss distribution was found to be minor. \cite{hertl21a, martin-barrios2022} Our findings also make sense from a qualitative point of view, as the energy exchange between projectile and surface can be reasonably well approximated by a binary collision model in which the relative energy transfer scales with $m/M$ with $m$ and $M$ being the mass of the projectile and substrate atom, respectively.\cite{grimmelmann80, hertl21b} The mass ratio for H/Ge is $\sim 1/73$, so that one can immediately see that the influence of energy exchange directly to the lattice is minor compared to the non-adiabatic effects that drive the energy transfer for scenarios $E_i > E_\text{gap}$.

\section{Conclusions}
We have introduced an electronic structure model parametrized for hydrogen atom scattering on a Ge(111)$c(2\times8)$ surface. The model considers a single adsorbate state that can switch between a neutral (repulsive) and a negatively charged (attractive) interaction potential upon charge transfer with the surface. Combined with a modified discretization scheme that introduces a bandgap into the spectrum of the electronic states, we constructed a one-dimensional Haldane-Anderson Hamiltonian model suitable for MQCD simulations of gas-surface scattering. We employed mean-field Ehrenfest dynamics,  independent electron surface hopping, and hierarchical equation of motion quantum dynamics simulations to study nonadiabatic energy loss as a function of incidence kinetic energy. 

The IESH simulations were able to capture a key feature observed in hydrogen atom scattering experiments from Ge(111)$c(2\times8)$: For kinetic energies below the bandgap, the scattered hydrogen atoms do not show any energy loss due to electronic excitations into the Ge conduction band. Above the bandgap, the scattered hydrogen atoms lose energy due to electronic transitions. On the other hand, mean field Ehrenfest dynamics of hydrogen atom scattering exhibit nonadiabatic energy loss under all conditions---an unphysical artefact caused by the mean-field nature of the method. This behavior is in contradiction to experimental observations. We, therefore, conclude that mean-field dynamics are not suitable to describe nonadiabatic effects during reactive scattering at semiconductor surfaces. We further derived kinetic energy loss distributions from IESH simulations for three experimentally accessible incidence energies and, given the simplicity of our model, achieved good agreement with the experiment{---}in particular for low initial kinetic energies. Our IESH simulations are also in good agreement with numerically exact HEOM simulations on the same model.


The current model is limited in the description of the electronic structure. With more and more advanced input on the excited electronic states through electronic structure calculations, a more accurate and higher-dimensional model may be parametrized that is better suited to describe the previously reported site specificity of nonadiabatic effects on germanium and energy loss at the highest measured kinetic energies.~\cite{zhuMechanisticInsightsNonadiabatic2024} 
This work outlines how diabatic energy curves and their couplings from first-principles calculations can be used to construct a physically suitable Hamiltonian. In addition to the reported MQCD simulations on this model, the HEOM results form a firm benchmark for the future development of improved MQCD methods.

\section{Acknowledgments}
    Funding is acknowledged from the UKRI Future Leaders Fellowship programme (MR/X023109/1), a UKRI frontier research grant (EP/X014088/1), a MSCA postdoctoral fellowship (EP/Z001498/1), the EPSRC Doctoral Training Partnership at the University of Warwick, and an Erasmus+ Traineeship mobility grant (2023-1-IT02-KA131-HED-000131307). High-performance computing resources were provided via the Scientific Computing Research Technology Platform of the University of Warwick, the EPSRC-funded Materials Chemistry Consortium (EP/R029431/1, EP/X035859/1), and the UK Car-Parrinello consortium (EP/X035891/1) for the ARCHER2 UK National Supercomputing Service, and the EPSRC-funded HPC Midlands+ computing centre for access to Sulis (EP/P020232/1).
    R.J.P. thanks the Alexander von Humboldt Foundation for the award of a Research Fellowship. The authors acknowledge the support by the state of Baden-W\"urttemberg through bwHPC and the DFG through Grant No. INST 40/575-1 FUGG (JUSTUS 2 cluster).

\section{Code Availability}
NQCDynamics.jl is open-source and available at \url{https://github.com/NQCD/NQCDynamics.jl}.

\providecommand{\noopsort}[1]{}\providecommand{\singleletter}[1]{#1}%

\end{document}


\title{Supporting Information for ``A Haldane-Anderson Hamiltonian Model for Hyperthermal Hydrogen Scattering from a Semiconductor Surface''}
\author{Xuexun Lu}
\affiliation{Department of Chemistry, University of Warwick, Gibbet Hill Road, CV4 7AL, Coventry, UK}
\author{Nils Hertl}%
\affiliation{Department of Chemistry, University of Warwick, Gibbet Hill Road, CV4 7AL, Coventry, UK}%
\affiliation{Department of Physics, University of Warwick, Gibbet Hill Road, CV4 7AL, Coventry, UK}
\author{Sara Oregioni}
\affiliation{Department of Chemistry, Universit{\`a} degli Studi di Milano, Via Golgi 19, 20133 Milano, Italy}
 \author{Riley Preston}
  \affiliation{Institute of Physics, University of Freiburg, 79104 Freiburg, Germany}
\author{Samuel L. Rudge}
 \affiliation{Institute of Physics, University of Freiburg, 79104 Freiburg, Germany}
 \author{Michael Thoss}
 \affiliation{Institute of Physics, University of Freiburg, 79104 Freiburg, Germany}
\author{Rocco Martinazzo}
\affiliation{Department of Chemistry, Universit{\`a} degli Studi di Milano, Via Golgi 19, 20133 Milano, Italy}
\affiliation{Instituto di Scienze e Tecnologie Molecolari, CNR, via Golgi 19, 20133 Milano, Italy}
\author{Reinhard J. Maurer}
\email{r.maurer@warwick.ac.uk}
\affiliation{Department of Chemistry, University of Warwick, Gibbet Hill Road, CV4 7AL, Coventry, UK}%
\affiliation{Department of Physics, University of Warwick, Gibbet Hill Road, CV4 7AL, Coventry, UK}
\affiliation{ University of Vienna, Faculty of Physics, Kolingasse 14{--}16, 1090 Vienna, Austria}
\maketitle

\date{\today}

\section{Optimisation of system-bath coupling Hamiltonian}
 The system-bath Hamiltonian which describes the proposed hydrogen atom coupling with an electronic band is defined as
\begin{align}
    H_{\text{sys-bath}} = U_0(x) + 
\begin{pmatrix}
h(x) & V_{1}(x) & V_{2}(x) & \cdots & V_{n}(x) \\
V_{1}(x) & \epsilon_1 & 0 & \cdots & 0 \\
V_{2}(x) & 0 & \epsilon_2 & \cdots & 0 \\
\vdots & \vdots & \vdots & \ddots & \vdots \\
V_{n}(x) & 0 & 0 & \cdots & \epsilon_n \\
\end{pmatrix}. \label{eq:Hamiltonian-matrix-NA}
\end{align}
To align the adiabatic ground-state potential energy surface derived from \( H_{\text{sys-bath}} \) with ground-state energies obtained from first-principles (DFT) calculations, we first fit a polynomial \( P_{\text{gr}} \) to DFT ground-state energies evaluated at a range of positions. This polynomial serves as a smooth reference for the target energy landscape.

We then seek a reasonable set of model parameters—as listed in TABLE I and the rescaling factor \( \bar{a} \)—by solving the following minimization problem:
\[
\text{minimize } \mathcal{L}_2(\mathbf{x})
\]
where
\[
\mathcal{L}_2(\mathbf{x}) := \frac{1}{N_{\text{opt}}} \sum_{i=1}^{N_{\text{opt}}} \left\| \lambda_{\text{min}}\left( H_{\text{sys-bath}}(x_i) \right) - P_{\text{gr}}(x_i) \right\|^2_2
\]
The sampled positions \( \mathbf{x} \) are taken uniformly from the interval \( [x_{\text{min}}, x_{\text{max}}]_{\text{ab}} \) corresponding to the ab initio evaluated position range. Here, \( \lambda_{\text{min}} \) denotes the smallest eigenvalue of the system-bath Hamiltonian at a given position, corresponding to the adiabatic ground-state energy.

\begin{figure}
    \centering
    \includegraphics[width=3.3in]{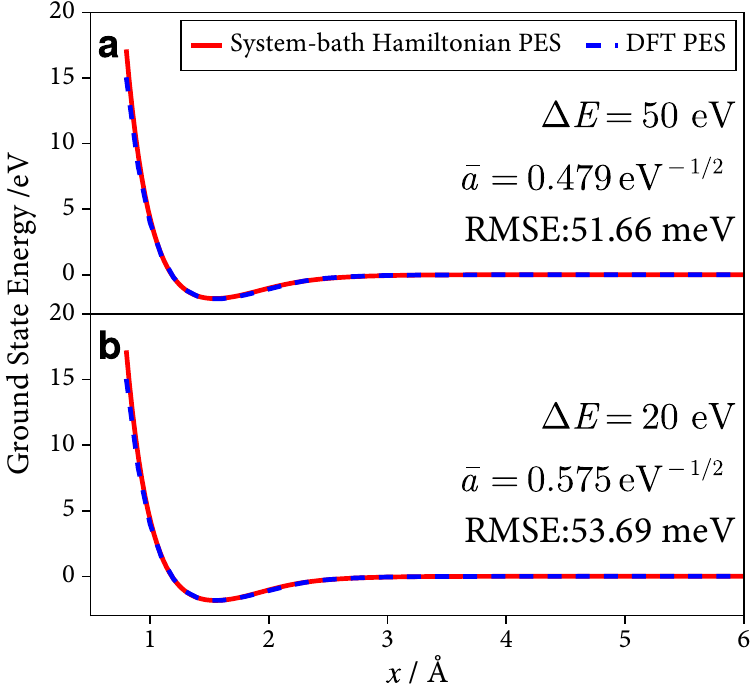}
    \caption{Comparison of the ground state potential energy surface (PES) from DFT (blue dashed) and a system-bath Hamiltonian, $H_{\text{sys-bath}}$ (solid red), for 2 different electronic bandwidths $\Delta E$ = 50\,eV (Panel \textbf{a}) and 20\,eV (Panel \textbf{b}), optimized via differential evolution\cite{stornDifferentialEvolutionSimple1997}. The corresponding rescaling factors $\bar{a}$ are 2.5 a.u. (0.479 $\text{eV}^{-1/2}$) and 3.0 a.u. (0.575 $\text{eV}^{-1/2}$), respectively. The root mean square errors (RMSE) between $H_{\text{sys-bath}}$ and DFT, evaluated from 1 to 6\,\AA, are $51.66$\,meV (Panel \textbf{a}) and $53.69$\,meV (Panel \textbf{b}).}
    \label{fig:width-50-20}
\end{figure}

The numerical gradient method (finite differencing of the loss function \(\mathcal{L}_2\)) was not employed due to the complexity of optimizing 14 parameters. Likewise, automatic differentiation was avoided because the involvement of eigenvalue decomposition complicates the exact determination of gradients. Instead, differential evolution \cite{stornDifferentialEvolutionSimple1997}, an iteratively direct searching method, can solve this multidimensional non-differentiable problem by selecting reasonable initial searching intervals. \cref{fig:width-50-20} illustrates the comparison between the optimized adiabatic ground-state potential energy surface from \( H_{\text{sys-bath}} \) and the DFT-based potential energy surface, using differential evolution as implemented in \texttt{BlackBoxOptim.jl}~\cite{vaibhav_kumar_dixit_2023_7738525} (version 0.6.3).

A different discretized constant bath width does not require refitting all parameters in \cref{eq:Hamiltonian-matrix-NA}. Using the diabatic state parameters listed in TABLE I, only the rescaling factor $\bar{a}$ needs to be adjusted for a new effective bandwidth. For example, comparing $\Delta E = 20$\,eV (\cref{fig:width-50-20}\textbf{a}) with the default setting $\Delta E = 50$\,eV (\cref{fig:width-50-20}\textbf{b}) used in the main text, we find that increasing $\bar{a}$ from 2.5 to 3.0 a.u., while keeping all other parameters fixed, yields a ground-state potential energy surface (PES) of $H_{\text{sys-bath}}$ that closely matches the DFT reference. This observation empirically supports the robustness of the introduced diabatic model and its associated parameters.

\section{ Modified trapezoidal discretization scheme}
\label{sec:GappedTrapezoidalRule}
A more intuitive method, \textit{Gapped Trapezoidal Rule} splits the Germanium band into two evenly discretised continuums, is given as 
\begin{equation}
    \label{eq:GapTrapezoidalRule-states}
    \epsilon_k = \begin{cases}E_{\text{F}} - \Delta E + (k-1) \times \frac{\Delta E - E_{\text{gap}}}{M} &\text { if } k \leq M/2 \\
   E_{\text{F}} + \frac{E_{\text{gap}}}{2} +  (k-M/2 -1) \times \frac{\Delta E - E_{\text{gap}}}{M}&\text { otherwise }\end{cases}
\end{equation}
attached with constant coupling weights
\begin{equation}
    \label{eq:GapTrapezoidalRule-weights}
    \omega_k = \frac{\Delta E - E_{\text{gap}}}{M}. 
\end{equation}

\begin{figure}
    \centering
    \includegraphics[width=3.3in]{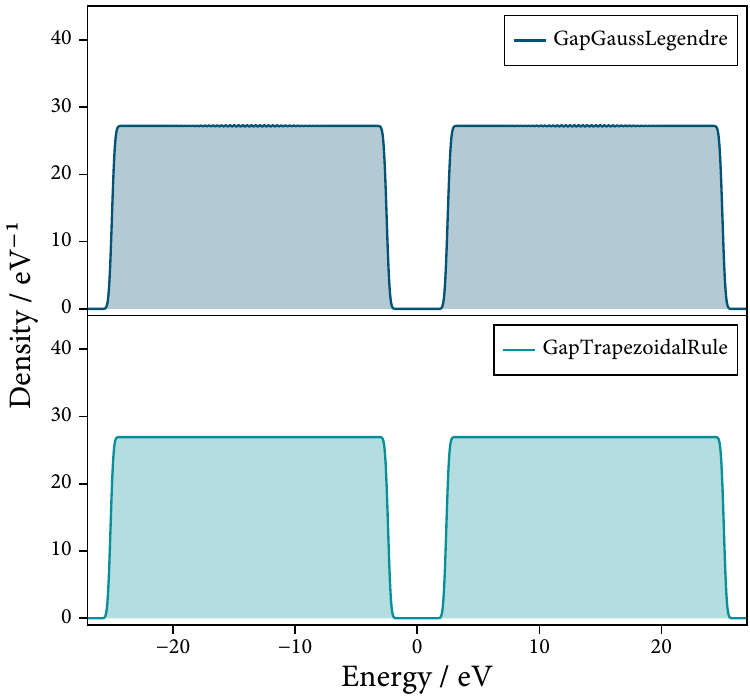}
    \caption{Density of states of two wide-band-limit baths discretised by Gapped Trapezoidal Rule and Gapped Gauss Legendre method respectively. Both of them have band gap $E_{\text{gap}}$ = 5 eV, number of discretisation $M = 500$, Fermi level $E_{\text{F}} = 0$ eV and bandwidth $\Delta E = $50 eV. The standard deviation is chosen as $\delta_{\text{sd}} = 0.25$.}
    \label{fig:bath-DOS}
\end{figure}

Both of them could produce an exact same uniform distributed bath's DOS while having a gap as FIG. \ref{fig:bath-DOS}. It can be computed by a summation of a series of Gaussian densities $\mathcal{N}(\epsilon_i, \delta_\text{sd})$ centered at knots $\epsilon_i$ with a given broadening deviation $\delta_{\text{sd}}$.
\begin{equation}
\rho_{\text{bath}} \approx \sum_k \omega_k \cdot\mathcal{N}(\epsilon_k, \delta_{\text{sd}})    
\end{equation}

However, the key difference of these two methods is Gapped Gauss-Legendre has more dense discretised energies around the boundaries and relatively sparse in the heart of two continuums. This feature allows Gapped Gauss-Legendre to have a easier electron excitation from HOMO's neighbor to LUMO's neighbor than the Gapped Trapezodial Rule with the same number of knots.

\section{Model Validation and Convergence Behaviour}

We investigated the effect of the selected electronic bath bandwidth $\Delta E$ and its number of states $M$ on the inelastic scattering energy loss of H atoms within our model.
\begin{figure}[H]
    \centering
    \includegraphics[width=3.3in]{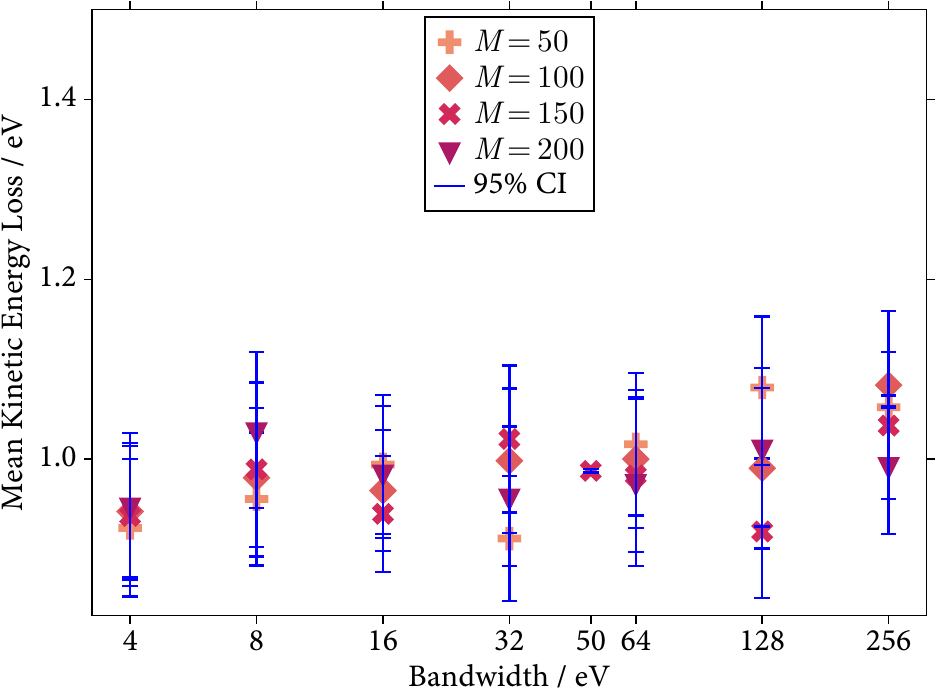}
    \caption{Convergence tests with respect to bandwidth $\Delta E$ and number of states $M$, discretized using the Gapped Gauss-Legendre method. Each marker represents the averaged inelastic kinetic energy loss from IESH simulations with incident energy $E_i = 1.92$ eV, released at 5~\AA. The marker at $\Delta E = 50$ eV and $M = 150$ represents the mean inelastic energy loss acquired from 180{,}899 trajectories. All other markers represent calculation sets consisting of over 500 trajectories.}
    \label{fig:bath-convergence}
\end{figure}

In \cref{fig:bath-convergence}, the mean kinetic energy loss exhibits convergence behavior across the scanned bandwidth $\Delta E$. Through this entire range of bandwidths, the values for the mean kinetic energy loss scatters between 0.9\,eV and 1.1\,eV which can be explained with the numerical uncertainty quantified via the 95\% confidence interval. Moreover, the mean kinetic energy loss is also very insensitive to the number of discretized states used in the IESH simulations. The spread of these values can be again explained with numerical uncertainty as the confidence intervals of the different data sets overlap. The insensitivity of the simulations with respect to the investigated parameter spaces showcases the robustness of our model which we further confirmed with a simulation set with $\Delta E = 50$\,eV and $M=150$ states in which we used about 180{,}000 inelastic simulations to reduce numerically uncertainty. We assign the robustness of our model with respect to $\Delta E$ and $M$ with the parameterization of our model. The probability for electronic excitations within the IESH simulations is determined by the energetic difference between the adiabatic ground state PES and the first couple adiabatic excited states close to the surface. This energetic difference is in turn governed by $h(x)$ and $A(x)$ which are constant at close distances to the surface{---}see FIG. 1. Consequently, as long as this crossing point between the adiabatic ground state and the adiabatic first excited state is properly modeled at a ground-state energy of about 0.49\,eV, the resulting energy loss of the simulations will be rather independent of the chosen parameters $\Delta E$ and $M$. 
\subsection{Sticking Probability}
\begin{figure}[H]
    \centering
    \includegraphics[width=3.3in]{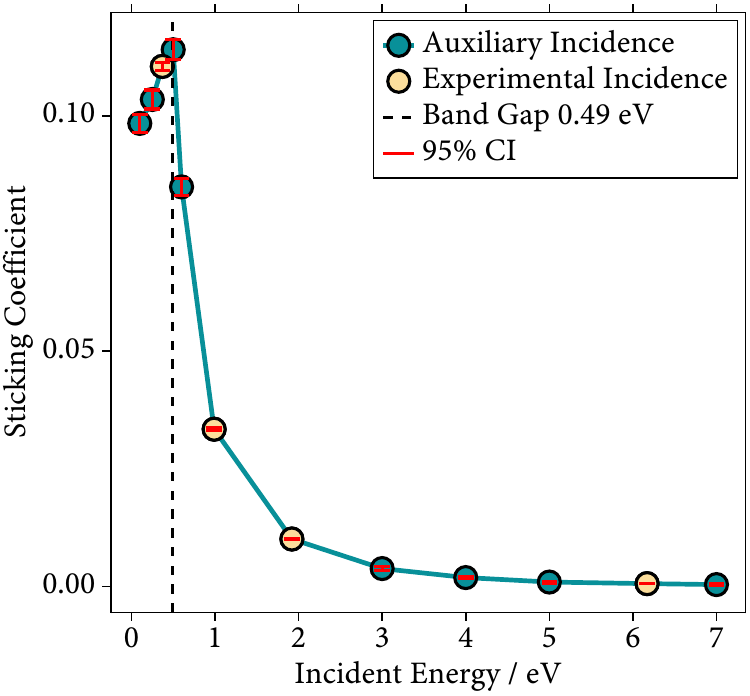}
    \caption{Sticking coefficient extracted from IESH simulations for various incidence energies. The yellow colored markers mark simulations with initial kinetic energies realizable in the experiments,\cite{kruger2023} i.e., 0.37, 0.99, 1.92, 6.17 eV. The teal colored markers represent simulations performed with other incidence energies to obtain a bigger picture and the following values were chosen: 0.1, 0.25, 0.5, 0.6, 3, 4, 5, 7 eV.}

    \label{fig:IESH-sticking}
\end{figure}
In addition to kinetic energy loss distributions, we report sticking coefficients which are experimentally inaccessible. To study the impact of the bandgap, we conduct IESH simulations over a broad range of initial kinetic energies ranging from 0.1\,eV to 7\,eV. The resulting sticking coefficients $S$ show finite sticking probability throughout the entire range of investigated kinetic energies with a distinct energy dependence of the sticking coefficients (\cref{fig:IESH-sticking}). In the energy regime $E_i < E_\text{gap}$, sticking increases, whereas for $E_i > E_\text{gap}$, sticking decreases monotonically.  Note that this is not in contrast to our finding that no  \textit{scattered} particles exhibit any energy loss in the energy regime $E_i < E_\text{gap}$. This finding can be reasoned in the following way: For very small incidence energies, the hydrogen atom does not make it to a height where the adiabatic ground and the adiabatic excited states start to cross and the probability for electron hops{---}the only energy transfer channel present in our simulations{---}is small but finite. In these slow-incidence cases, the hydrogen atom will not escape from the surface once an electron hop takes place and the effective potential can not return back to the adiabatic ground state. If the initial kinetic energy is approaching the value of the bandgap, the hydrogen atom comes closer to the region where the different adiabatic potential energy surfaces intersect and the likelihood for nonadiabatic energy transfer and thus sticking increases. Once the initial kinetic energy starts to exceed the bandgap, the hydrogen atoms have a larger reservoir of kinetic energy that can be dissipated by electronic excitations and therefore can escape the binding well of the substrate more easily. In addition, the higher velocity of the hydrogen atoms make them spend a smaller amount of time in the region where nonadiabatic effects are most likely to occur. Overall, for the regime $E_i > E_\text{gap}$, the energy dependence of sticking we find in our simulations agrees with earlier reports by Trail \textit{et al}.\cite{trail03} for sticking of hydrogen on Cu(111). In this study, the ability of hydrogen atoms to stick to the metal surface is investigated using a one-dimensional Newns-Anderson Hamiltonian, where electron-hole pair excitations are the only energy transfer channel{---}just as in our simulations{---}but are simulated with the molecular dynamics with electronic friction method.~\cite{head-gordon95} The distinct difference between H/Cu(111) and H/Ge(111)$c(2\times8)$ is therefore the presence of the bandgap, which acts similarly to an electronic barrier for the sticking process. Again, we stress that we only consider the nonadiabatic energy transfer channel here and once energy exchange with phonons is introduced, the overall energy dependence of sticking might change once the transition to full-dimensional potentials is made. 

\section{Bader charge}
\begin{figure}[H]
    \centering
    \includegraphics[width=4in]{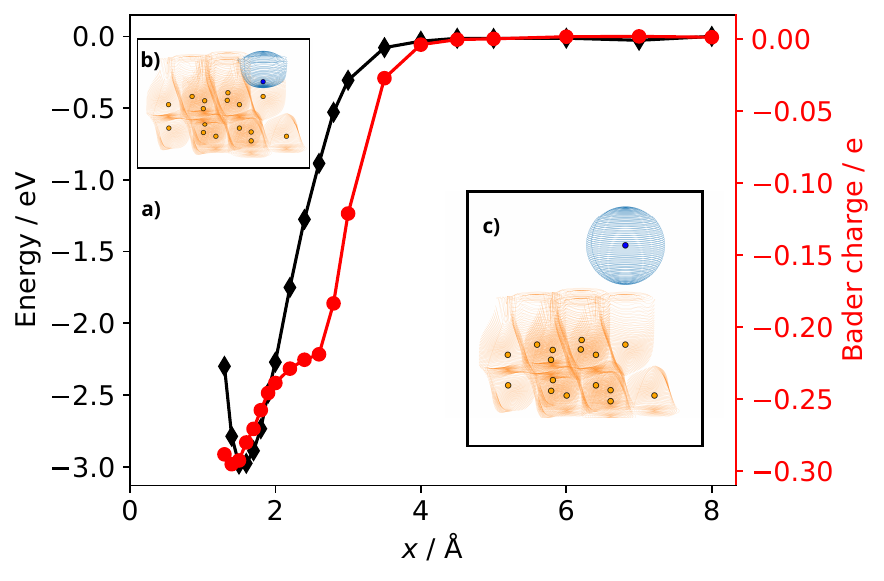}
    \caption{Adiabatic DFT energies (black diamonds) and the net Bader charge of  hydrogen (red circles) as a function of vertical distance to the Ge(111)$c(2\times8)$ surface, $x$. The 1D grid is shown in panel a), whereas panel b) and c) contain contour plots of the charges for the lowest and highest height value of $x$, respectively. The blue and orange contour lines indicate the surface of the Bader volumes for hydrogen and germanium, respectively. }
    \label{fig:Bader_charge}
\end{figure}

\section{Illustrative Figures}

\begin{figure}[H]
    \centering
    \includegraphics[width=3.3in]{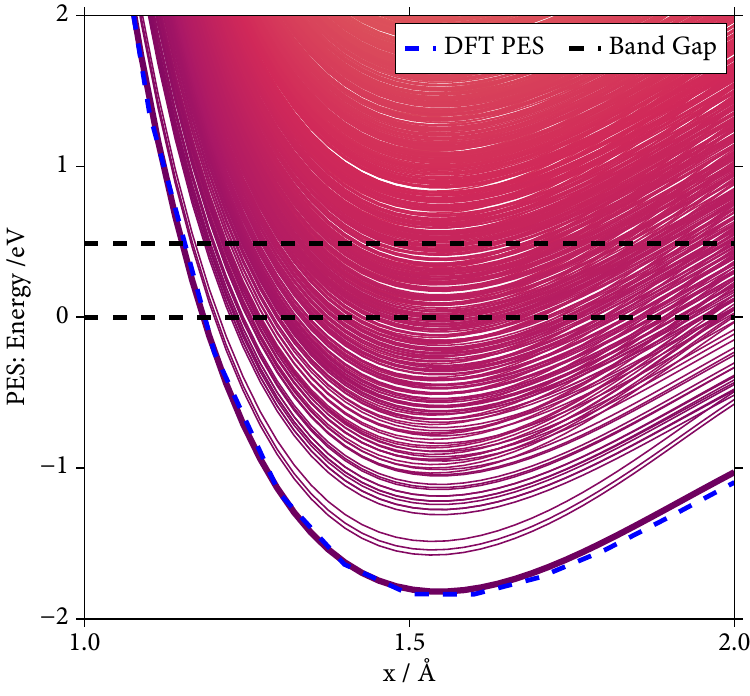}
    \caption{Adiabatic potential energy spectrum for the 2000 lowest energy many-electron states zoomed from 1 to 2\,\AA\, and -2 to 2\,eV (a zoomed-in version of Fig.\,2(b) in main text). The spectrum is produced with the gap-inserting Gauss-Legendre method using 150 states, 75 electrons and a bandwidth of $\Delta E =$ 50\,eV. The inserted band gap has width $E_{\text{gap}} = 0.49$\,eV. The Fermi level is $E_\text{F}= 0 $\,eV. The rescaling parameter for the couplings is  $\bar{a} = 0.479\,\text{eV}^{-1/2}$.}
    \label{fig:bath-convergence}
\end{figure}

\section{Exploring $\sigma$ in Wigner Distribution}
In this section, we assessed the robustness of the HEOM calculations on the initial conditions like the incidence energy $E_i$ and the spatial width of the Gaussian wavepacket $\sigma$ and complemented it with corresponding IESH calculations. As for the results shown in the results section of the main manuscript, the agreement between IESH and HEOM is good for a broad range of initial kinetic energies and wavepacket shapes.  

\begin{figure}[H]
    \centering
    \includegraphics[width=3.3in]{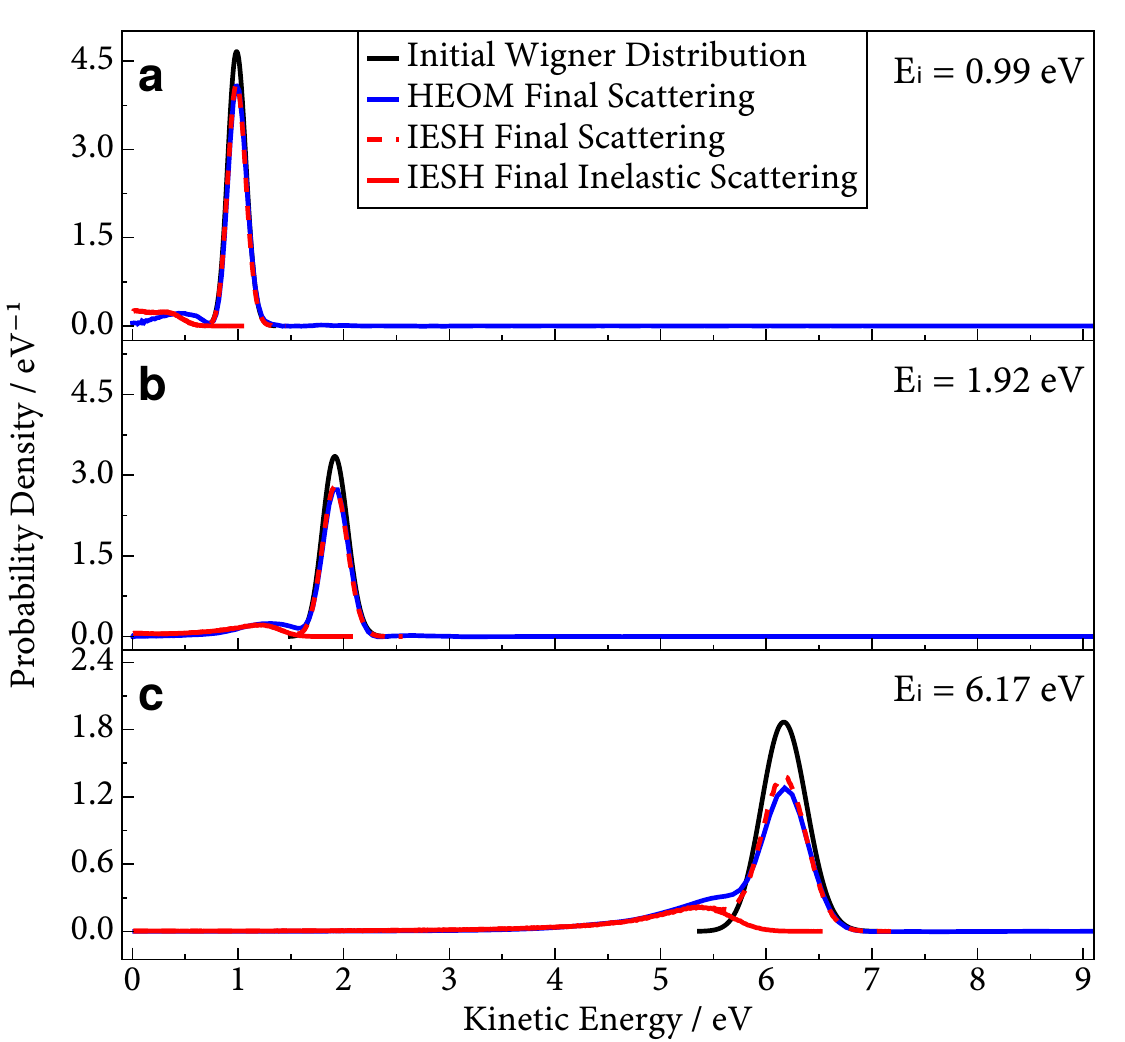}
    \caption{Probability density of the final kinetic energy for hydrogen atom scattered from Ge(111)$c(2\times8)$, simulated using IESH and HEOM. Results are shown for incidence energies $E_i = 0.99$~eV(\textbf{a}), 1.92~eV(\textbf{b}), and 6.17~eV(\textbf{c}) at $T = 300$~K. The initial states are the Wigner distribution with width $\sigma = 1.0$\,a.u (black line). The full scattering distribution (red dashed line), which includes both elastic and inelastic contributions, and the inelastic scattering distribution (red baseline) from IESH were constructed using a Gaussian kernel density estimate (KDE) with a standard deviation $\delta_{\rm sd} = 0.005$. The probabilities for inelastic scattering events are 0.114, 0.169, and 0.264 for 0.99~eV, 1.92~eV, and 6.17~eV, respectively.
    }
    \label{fig:IESH-HEOM-1.0}
\end{figure}

\begin{figure}[H]
    \centering
    \includegraphics[width=3.3in]{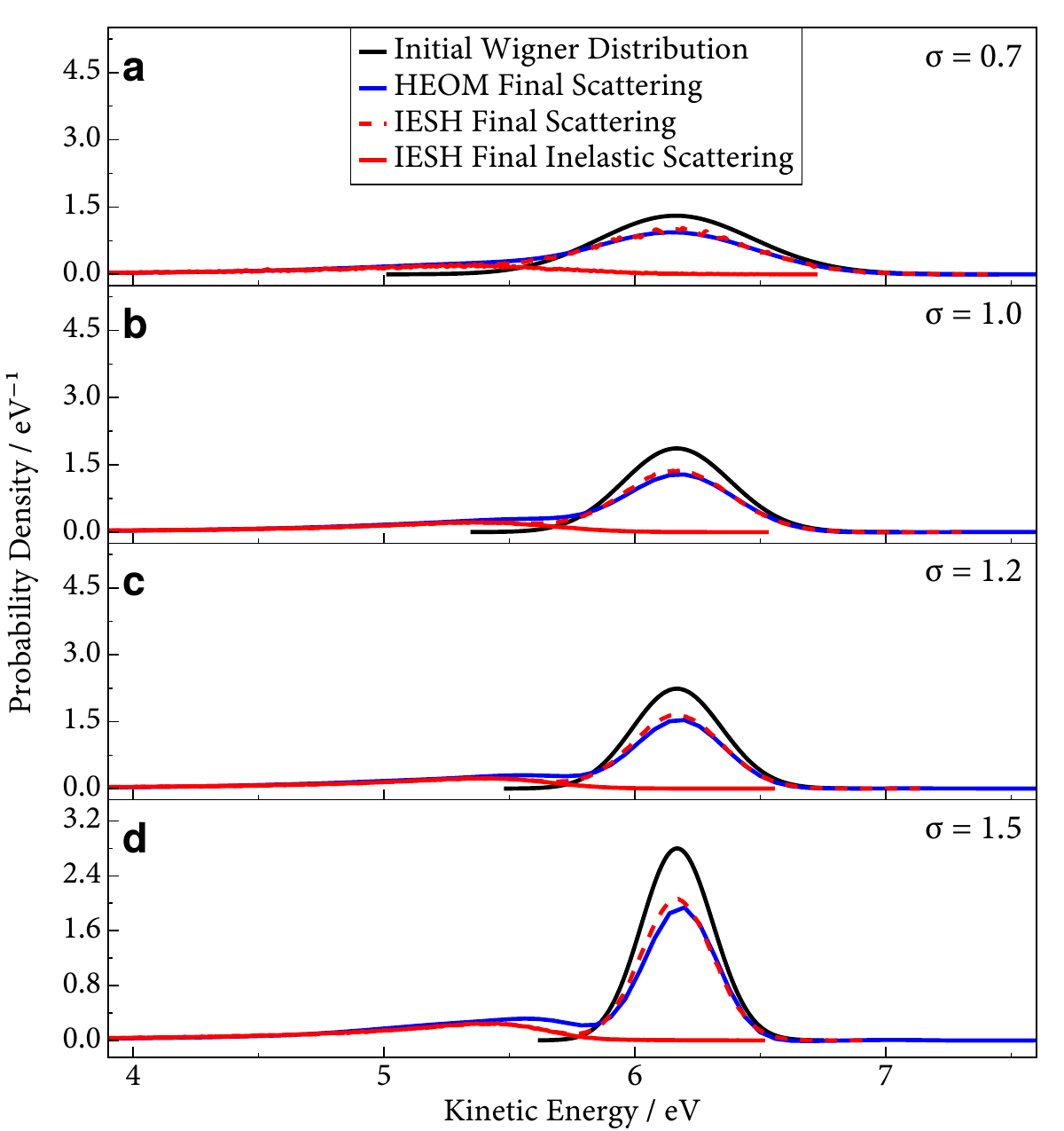}
    \caption{Probability density of the final kinetic energy for hydrogen atom scattered from Ge(111)$c(2\times8)$ with Wigner distributed initial knetic energies centered at 6.17\,eV, simulated using IESH and HEOM. Results are shown for Wigner distribution widths $\sigma = 0.7$\,a.u.(\textbf{a}), 1.0\,a.u.(\textbf{b}), 1.2\,a.u.(\textbf{c}), and 1.5\,a.u.(\textbf{d}) at $T = 300$~K. The initial states are the Wigner distributions centered at $E_i = 6.17$\,eV (black line). The full scattering distribution (red dashed line), which includes both elastic and inelastic contributions, and the inelastic scattering distribution (red baseline) from IESH were constructed using a Gaussian kernel density estimate (KDE) with a standard deviation $\delta_{\rm sd} = 0.005$. The probabilities for inelastic scattering events are around 0.2666, for all 4 selected $\sigma$. 
    }
    \label{fig:IESH-HEOM-1.0}
\end{figure}

\section{HEOM calculations with a phononic bath}
We performed HEOM calculations for H atom scattering from $c(2\times8)$ while including an additional phononic bath in the surface along with the already included fermionic bath. The theoretical methodology regarding the inclusion of both fermionic and phononic baths in HEOM calculations has been discussed in detail by Ke \textit{et al.}\cite{ke2022}. $\lambda$ is the strength of the coupling between the projectile and the phonons. We screened the role of phonons for several coupling strengths and found the effect of the phonon bath to be almost negligible. This result is consistent with previous observations made for H/W(110), where phonons have been modeled explicitly and with a generalized Langevin oscillator model.\cite{martin-barrios2022}
\begin{figure}[H]
    \centering
    \includegraphics[width=3.3in]{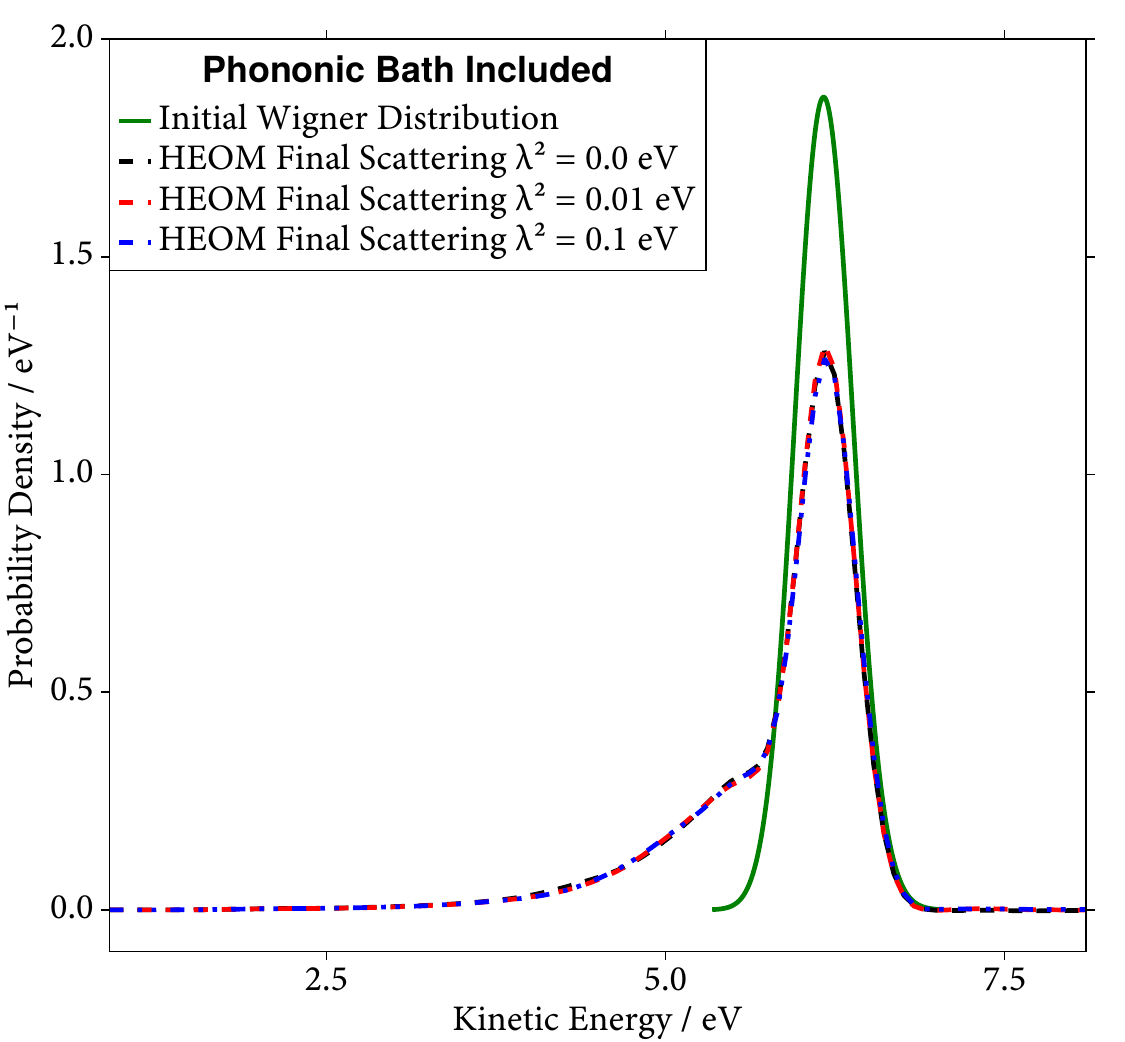}
    \caption{Probability density of the final kinetic energy for hydrogen atom scattered from Ge(111)$c(2\times8)$ simulated by HEOM with a phononic bath applied on top of the electronic bath. The green solid line shows the Wigner distributed initial state centered at kinetic energy 6.17\,eV and width $\sigma = 0.4$\,a.u.. The dashed curves show the HEOM final scattering distributions under the included phononic bath with three coupling strengths $\lambda^2 = 0.0$\,eV (black), $0.01$\,eV (red) and $0.1$\,eV (blue). The other phonon bath parameters are $\Omega_c = 15$meV and $\alpha = 0.51$\AA$^{-1}$.
    }
    \label{fig:IESH-HEOM-1.0}
\end{figure}

\bibliography{Literature,HEOM_bibli}